\documentclass[aps, prl, superscriptaddress, reprint, twocolumn, a4, notitlepage]{revtex4-1}
\usepackage{graphicx}
\usepackage{amssymb}
\usepackage{amsmath,bm}
\usepackage{dcolumn}
\usepackage{mathrsfs}
\usepackage{physics}
\usepackage[utf8]{inputenc}
\usepackage{mathptmx}
\DeclareMathAlphabet{\mathcal}{OMS}{cmsy}{m}{n}
\newcommand{\widebar}[1]{\mkern 1.5mu\overline{\mkern-1.5mu#1\mkern-1.5mu}\mkern 1.5mu}

\usepackage{color}

\newcommand{\bk}{\mathbf{k}}
\newcommand{\bp}{\mathbf{p}}

\newcommand{\bq}{\mathbf{q}}

\newcommand{\edit}[1]{#1}

\begin{document}
\title{Efficient quadrature-squeezing from biexcitonic parametric gain in atomically thin semiconductors}
\date{\today}

\author{Emil V. Denning}
\affiliation{Nichtlineare Optik und Quantenelektronik, Institut f\"ur Theoretische Physik, Technische Universit\"at Berlin, 10623 Berlin, Germany}

\author{Andreas Knorr}
\affiliation{Nichtlineare Optik und Quantenelektronik, Institut f\"ur Theoretische Physik, Technische Universit\"at Berlin, 10623 Berlin, Germany}

\author{Florian Katsch}
\affiliation{Nichtlineare Optik und Quantenelektronik, Institut f\"ur Theoretische Physik, Technische Universit\"at Berlin, 10623 Berlin, Germany}

\author{Marten Richter}
\affiliation{Nichtlineare Optik und Quantenelektronik, Institut f\"ur Theoretische Physik, Technische Universit\"at Berlin, 10623 Berlin, Germany}

\date{\today}

\begin{abstract}
Modification of electromagnetic quantum fluctuations in the form of quadrature-squeezing is a central quantum resource, which can be generated from nonlinear optical processes. Such a process is facilitated by coherent two-photon excitation of the strongly bound biexciton in atomically thin semiconductors. We show theoretically that interfacing an atomically thin semiconductor with an optical cavity allows to harness this two-photon resonance and use the biexcitonic parametric gain to generate squeezed light with input power an order of magnitude below current state-of-the-art devices with conventional third-order nonlinear materials that rely on far off-resonant nonlinearities. Furthermore, the squeezing bandwidth is found to be in the range of several meV. These results identify atomically thin semiconductors as a promising candidate for on-chip squeezed-light sources.
\end{abstract}
\maketitle


{\it Introduction.\textemdash}
Quadrature-squeezed light is important for many quantum-technological applications, e.g. metrology~\cite{caves1981quantum, taylor2013biological, aasi2013enhanced}, computing~\cite{lloyd1999quantum, cerf2001quantum,menicucci2006universal, menicucci2014fault}, communication~\cite{hillery2000quantum,cerf2001quantum} and simulation~\cite{huh2015boson,sparrow2018simulating,banchi2020molecular}. Since its first experimental realization using four-wave mixing in an atomic beam~\cite{slusher1985observation}, quadrature-squeezed light has been demonstrated in many material platforms, such as second-order nonlinear crystals in free space~\cite{wu1986generation,breitenbach1997measurement,vahlbruch2008observation,vahlbruch2016detection} and on integrated chips~\cite{furst2011quantum,harder2016single,lenzini2018integrated,otterpohl2019squeezed}, third-order nonlinearities in optical fibers~\cite{shelby1986broad,rosenbluh1991squeezed,bergman1991squeezing,finger2015raman} and on integrated chips~\cite{dutt2015chip,zhao2020near,vaidya2020broadband,zhang2021squeezed}, single-emitter resonance flourescence~\cite{schulte2015quadrature} and excitons in semiconductors~\cite{fox1995squeezed,karr2004squeezing,boulier2014polariton}; for a comprehensive review, see Ref.~\onlinecite{andersen201630}.

Quadrature squeezing is canonically described through the operator $\exp[(za^2 - za^{\dagger 2})/2]$, which reduces the in-phase quadrature noise of a single mode with photon annihilation operator $a$ by an amount of $\exp(-z)$~\cite{gardiner2004quantum}. Thus, pairwise photon creation, also known as \emph{parametric gain}, generates quadrature squeezing. Coherent excitation of the Coulomb-bound biexciton in semiconductors enables strong resonant enhancement of pairwise creation of energy quanta~\cite{lovering1992resonant,brunner1994sharp,hassan1994direct}, which can provide parametric gain for quadrature squeezing~\cite{shimano2002efficient}. This efficient two-photon resonance is absent in conventional off-resonant third-order nonlinear materials such as $\mathrm{Si_3N_4}$~\cite{dutt2015chip,zhao2020near,vaidya2020broadband,zhang2021squeezed}. Atomically thin semiconductors are particularly interesting in this context, because of their exceptionally strong Coulomb interaction~\cite{shahnazaryan2017exciton,katsch2018theory,stepanov2021exciton} and thus strongly bound biexciton~\cite{you2015observation,steinhoff2020dynamical}, owing to reduced dimensionality and dielectric screening~\cite{latini2015excitons}. Furthermore, polaritonic microcavities with atomically thin semiconductors have already been experimentally demonstrated on a photonic chip~\cite{liu2015strong,dufferwiel2015exciton,sidler2017fermi,anton2021bosonic,stepanov2021exciton}.

\begin{figure}
\centering
  \includegraphics[width=\columnwidth]{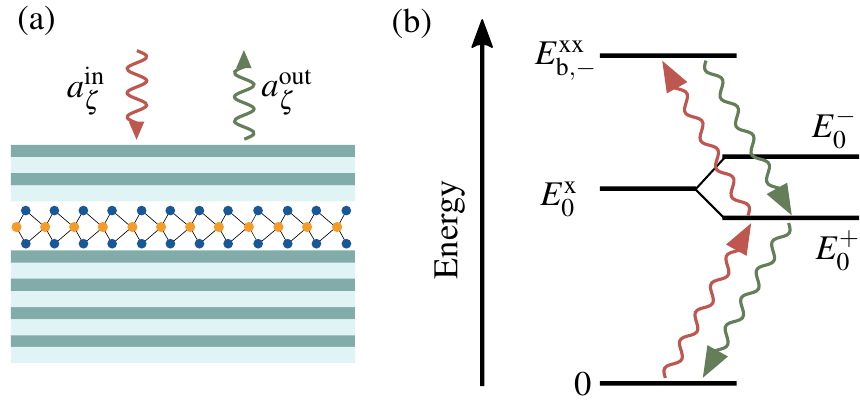}
    \caption{(a) Atomically thin semiconductor placed in a one-sided planar cavity driven by an optical input field $a^{\rm in}$. (b) Polariton energy bands with illustration of lower polaritons generated by the input field. (c) Two optically generated lower polaritons can form a bound biexciton via the Coulomb interaction ($\widebar{W}^{-}_{\rm b}$). When the polariton pair energy ($2E_0^-$) matches the bound biexciton ($E^{\rm xx}_{\rm b,-}$) the process is resonantly enhanced. (d) Bound biexcitons can provide parametric gain by breaking into correlated polariton pairs, which are outcoupled from the cavity as squeezed light.}
    \label{fig:1}
\end{figure}

In this Letter, we theoretically demonstrate that the biexciton allows generation of broadband quadrature-squeezed light on a photonic chip with very low input power (1-10 mW) an order of magnitude below state-of-the-art devices with conventional third-order nonlinearities~\cite{dutt2015chip,zhao2020near,vaidya2020broadband,zhang2021squeezed}.
\edit{We consider a laser-driven planar microcavity coupled to an atomically thin semiconductor [Fig.~\ref{fig:1}(a)-(b)], where two optically generated polaritons are converted into a bound biexciton via the many-body Coulomb interaction [Fig~\ref{fig:1}(c)], not present in typical atomic level schemes. When the energy of a lower polariton pair $2E_0^-$ matches that of the bound biexciton $E^{\rm xx}_{\rm b,-}$, the process is strongly resonant, and coherent biexcitons are efficiently excited. The generated biexcitons drive the polariton field by spontaneously breaking into pairs, thus providing parametric gain and squeezing [Fig.~\ref{fig:1}(d)].}


The analysis of quadrature squeezing in such systems faces two main challenges: first, accounting for the strong correlations generated predominantly by the Coulomb interaction;
second, the need for spectral resolution of the squeezing---the key observable in homodyne detection~\cite{collett1987quantum,carmichael1987spectrum}---which requires an evaluation of multitime correlation functions. Even though excitonic many-body correlation effects have been studied extensively in semiconductors~\cite{axt1994dynamics,lindberg1994chi,ostreich1995exciton,schafer1996femtosecond,savasta1996quantum,savasta1999hyper,kira1999quantum,kwong2001third,takayama2002T,savasta2003many,schumacher2005coherent,schumacher2006coherent,portolan2008dynamics,katsch2019theory,katsch2020exciton,katsch2020optical}, all existing theories of squeezed-light generation in polaritonic microcavities that include spectral resolution are based on mean-field theory and omit Coulomb many-body correlations beyond the Hartree-Fock level~\cite{eleuch1999cavity,messin1999squeezed,schwendimann2003statistics,quattropani2005polariton,romanelli2010two} or considered a phenomenological 1D model~\cite{oka2011light}.

\begin{figure}
\centering
  \includegraphics[width=\columnwidth]{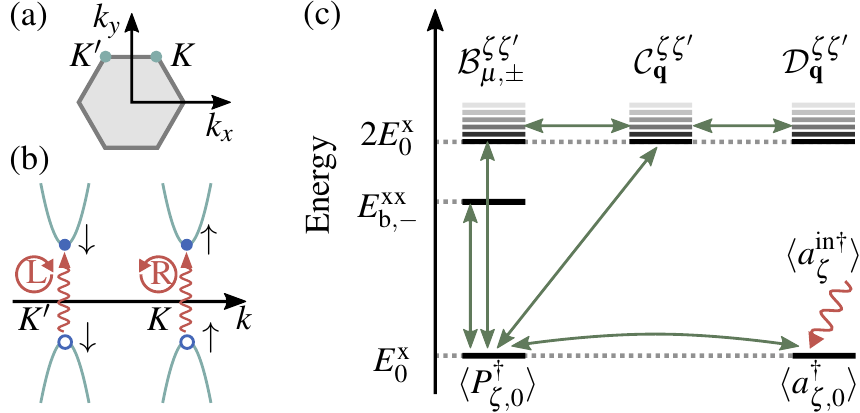}
    \caption{
    (a) 2D hexagonal Brillouin zone of atomically thin transition-metal dichalcogenide semiconductors with symmetry points $K$ and $K'$, where direct exciton transitions occur. (b) Circular optical selection rules for the transitions. (c) Energies of single-exciton and single-photon expectation values and multiparticle correlations (here indicated for $E^{\rm p}_0=E^{\rm x}_0$). The arrows visualize the coupling of expectation values in the dynamical evolution.}
    \label{fig:2}
\end{figure}



{\it Theory.\textemdash} The total Hamiltonian of the system is $H=H_0 + H_{\rm C}$, where $H_0$ describes free electrons, holes and photons and their coupling, and $H_{\rm C}$ describes Coulomb interactions; external driving is introduced through input-output formalism (see Supplementary Material~\cite{suppmat}). The bosonic photon annihilation (creation) operators $a_{\sigma \bq}^{(\dagger)}$, with polarisation $\sigma$ and in-plane momentum $\bq$ describe the electromagnetic field in the cavity. The fermionic annihilation (creation) operators $c_{\zeta\bk}^{(\dag)}$ and $v_{\zeta\bk}^{(\dag)}$ describe conduction and valence band electrons in the semiconductor, where the compound index $\zeta=(\xi,s)$ labels spin ($s$) and valley ($\xi$), and $\bk$ is the 2D wave vector.

For atomically thin transition-metal dichalcogenides, the energetically lowest optical transitions appear at the $K$ and $K'$ valleys in the Brillouin zone [cf. Fig.~\ref{fig:2}(a)]. Due to spin-orbit coupling, right/left-hand circularly polarised photons ($\sigma=\mathrm{R/L}$) can excite electron-hole pairs with \mbox{$\zeta=(K,\uparrow)$}/$\zeta=(K',\downarrow)$~\cite{mak2010atomically,splendiani2010emerging,kuc2011influence,xiao2012coupled,cao2012valley} [cf. Fig.~\ref{fig:2}(b)]. Thereby, photon polarisation in the circular basis is absorbed into the index $\zeta$ to label spin, valley and polarisation as $\zeta\in\{K, K'\}$.

The cavity photon energy is given by~\cite{skolnick1998strong,osgood2021dielectric} $E^{\rm p}_\bq = \hbar[\omega_{\rm p,0}^2 + (c q/\widebar{n})^2]^{1/2}$, where $\omega_{\rm p,0}$ is the resonance frequency of the cavity mode, $c$ is the vacuum speed of light and $\widebar{n}$ is the effective cavity refractive index. The cavity is taken to be one-sided with outcoupling rate $\gamma^{\rm p}$.

The $\bq=0$ mode of the cavity field is driven by coherent light with polarization vector $\bm{\lambda}_{\rm in}$ in the circular basis.
The quantity of interest is the quadrature operator of the cavity field at $\bq=0$, $X(\theta,t) = e^{i\theta} \bm{\lambda}_{\rm out}^{\rm T}\mathbf{a}_0^{\dagger}(t) + e^{-i\theta} \bm{\lambda}_{\rm out}^{\rm *T}\mathbf{a}_0(t)$, where $\mathbf{a}_0 = (a_{K,0},\: a_{K',0})^{\rm T}$ and $\bm{\lambda}_{\rm out}$ is the detected polarisation vector and the time argument $t$ denotes Heisenberg time evolution. While the absolute squeezing of the intracavity field can be calculated as the variance of $X(\theta,t)$, a more relevant measure is the squeezing of the outcoupled and thus detected field. This is characterized by the squeezing spectrum $\Lambda(\omega,\theta) = 2\sqrt{2\gamma^{\rm p}}\int_{0}^\infty\dd{\tau} \cos(\omega\tau)\ev*{:\delta X(\theta,\tau)\delta X(\theta,0):}$~\cite{carmichael1987spectrum}, where $\delta X(\theta,t) := X(\theta,t)-\ev*{X(\theta,t)}$. The symbols $::$ denote normal- and time-ordering, such that the time argument increase to the right in products $a^\dagger$ and to the left in products of $a$. The photocurrent noise spectrum of homodyne detection, normalised to the shot-noise level, is given by $1+\Lambda(\omega,\theta)$, where $\theta$ is the homodyne phase~\cite{carmichael1987spectrum}. We use the shorthand notation $\Lambda(\omega)$ to denote the squeezing spectrum at the optimal homodyne phase $\theta$ giving the lowest value of $\Lambda(\omega,\theta)$.

In the numerical calculations presented in this paper, we use strictly linear polarization, $\bm{\lambda}_{\rm in} = 2^{-1/2}[1,1]$, since this allows excitation of the bound biexciton~\cite{katsch2019theory}. For detection, we consider the co-polarized ($\bm{\lambda}_{\rm out}=\bm{\lambda}_{\rm in}$) and the cross-polarized ($\bm{\lambda}_{\rm out}=2^{-1/2}[1,-1]$) configurations.

\edit{Exciton creation operators are introduced by expanding electron-hole pairs on exciton wavefunctions as $P_{\zeta,\bq}^{n\dagger} = \sum_{\bk} \phi_{\bk}^nc_{\zeta\bk+\alpha\bq}^\dagger v_{\zeta\bk-\beta\bq}$, where $\phi_\bk^n$ is the momentum-space wavefunction of the $n$th exciton state} obtained from the Wannier equation~\cite{wannier1937structure,sham1966many,kira2006many} and $\alpha=m_{\rm e}/(m_{\rm e} + m_{\rm h}),\; \beta=m_{\rm h}/(m_{\rm e} + m_{\rm h})$ are coefficients defined from the electron ($m_{\rm e}$) and hole ($m_{\rm h}$) masses. \edit{The lowest-energy exciton ($n=\mathrm{1s}$) is separated from the next state by hundreds of meV~\cite{mak2010atomically,ramasubramaniam2012large}. Due to this large energy gap and assuming excitation in the vicinity of the 1s exciton energy, we truncate the electronic pair space to the 1s exciton subspace and omit the index $n$}. Within the effective mass approximation, the exciton energy is $E_{\bq}^{\rm x} = E_0^{\rm x} + \hbar^2 q^2/[2(m_{\rm e}+ m_{\rm h})]$, with $E_0^{\rm x}$ exciton energy for $\bq=0$.

To study the leading nonlinear response, we apply the dynamics-controlled truncation scheme~\cite{axt1994dynamics,lindberg1994chi,savasta1996quantum} to expand the equations of motion to third order in the driving field $a^{\rm in}_\zeta$, which corresponds to keeping only terms up to three normal-ordered electron-hole pair or photon operators. In the Supplementary Material, all details of the derivation are described: A closed set of equations is obtained for the zero-momentum exciton and photon expectation values $\ev*{a^\dagger_{\zeta, 0}}$ and $\ev*{P^\dagger_{\zeta,0}}$ and three types of correlations. Two-photon correlations are defined as $\mathcal{D}^{\zeta\zeta'}_\bq := \ev*{a_{\zeta\bq}^\dagger a_{\zeta'-\bq}^\dagger} - \ev*{a_{\zeta\bq}^\dagger}\!\!\ev*{a_{\zeta'-\bq}^\dagger}$.
Electron-hole-photon correlations $\ev*{a_{\zeta,\bq}^\dagger c^\dagger_{\zeta',\bk-\alpha\bq}v_{\zeta',\bk+\beta\bq}}^{\rm c} = \ev*{a_{\zeta,\bq}^\dagger c^\dagger_{\zeta',\bk-\alpha\bq}v_{\zeta',\bk+\beta\bq}} - \ev*{a_{\zeta,\bq}^\dagger}\!\!\ev*{ c^\dagger_{\zeta',\bk-\alpha\bq}v_{\zeta',\bk+\beta\bq}}$ \edit{are projected onto the 1s exciton} wavefunctions as $\mathcal{C}^{\zeta\zeta'}_\bq := \sum_\bk \phi_{\bk}\ev*{a_{\zeta,\bq}^\dagger c^\dagger_{\zeta',\bk-\alpha\bq}v_{\zeta',\bk+\beta\bq}}^{\rm c}$.

Two-pair correlations are defined as $\ev*{c^\dagger_{\zeta\bk+\bq}v_{\zeta\bk}c^\dagger_{\zeta'\bk'-\bq}v_{\zeta'\bk'}}^{\rm c} := \ev*{c^\dagger_{\zeta\bk+\bq}v_{\zeta\bk}c^\dagger_{\zeta'\bk'-\bq}v_{\zeta'\bk'}} - \ev*{c^\dagger_{\zeta\bk+\bq}v_{\zeta\bk}}\!\!\ev*{c^\dagger_{\zeta'\bk'-\bq}v_{\zeta'\bk'}} + \ev*{c^\dagger_{\zeta'\bk'-\bq}v_{\zeta\bk}}\!\!\ev*{c^\dagger_{\zeta\bk+\bq}v_{\zeta'\bk'}}$.
These are first \edit{projected} on the 1s-exciton wavefunction and then partitioned into singlet $(-)$ and triplet $(+)$ channels, defining the biexcitonic correlations $\tilde{\mathcal{B}}^{\zeta\zeta'}_{\bq,\pm}$ through the relation~\cite{schafer2013semiconductor}
\begin{align*}
&\frac{1}{2}\qty(\ev*{c_{\zeta\bk+\bq}^\dagger v_{\zeta\bk}
     c_{\zeta\bk'-\bq}^\dagger v_{\zeta'\bk'}}^{\rm c}
\pm
\ev*{c_{\zeta'\bk+\bq}^\dagger v_{\zeta\bk}
     c_{\zeta\bk'-\bq}^\dagger v_{\zeta'\bk'}}^{\rm c})
\\     &=:
\phi_{\bk + \beta\bq}^{*}
\phi_{\bk' - \beta\bq}^{*}
\tilde{\mathcal{B}}_{\bq,\pm}^{\zeta\zeta'}
\mp
\phi_{\alpha\bk + \beta(\bk'-\bq)}^{*}
\phi_{\beta(\bk+\bq) + \alpha\bk'}^{*}
\tilde{\mathcal{B}}_{\bk'-\bk-\bq,\pm}^{\zeta\zeta'}.
\end{align*}

These correlations have a more involved structure than exciton-photon and two-photon correlations because of the two possibilities of electron-hole pairing. We transform to a diagonalised biexcitonic basis via the wave functions $\Phi^{\pm}_{\mu\bq}$ with eigenenergies $E^{\rm xx}_{\mu,\pm}$ as $\tilde{\mathcal{B}}^{\zeta\zeta'}_{\bq,\pm}=\sum_\mu \Phi^{\pm}_{\mu\bq} \mathcal{B}^{\zeta\zeta'}_{\mu,\pm}$~\cite{katsch2019theory, suppmat}.
For the singlet channel, bound ($\mu=\mathrm{b},\:E_{\rm b,-}^{\mathrm{xx}}<2E_{0}^{\rm x}$) and unbound ($E_{\mu,-}^{{\rm xx}}>2E_{0}^{\rm x}$) solutions exist, whereas the triplet channel includes only unbound solutions~\cite{takayama2002T}. The unbound solutions constitute a correlated two-exciton scattering continuum.

Phonon-induced broadening of the excitonic and biexcitonic energies is introduced in the equations of motion through the complex energies $\tilde{E}^{\rm x}_\bq=E^{\rm x}_\bq+i\hbar\gamma^{\rm x},\; \tilde{E}^{\rm xx}_{\mu,\pm}=E^{\rm xx}_{\mu,\pm}+2i\hbar\gamma^{\rm x}$, with a self-consistent microscopically calculated $\gamma^{\rm x}$~\cite{selig2016excitonic,christiansen2017phonon,khatibi2018impact,brem2019intrinsic,lengers2020theory}, and we approximate the biexcitonic damping with $2\gamma^{\rm x}$~\cite{sieh1999coulomb,schumacher2005coherent,schumacher2006coherent}. Similarly, outcoupling from the cavity is introduced as $\tilde{E}^{\rm p}_\bq=E^{\rm p}_\bq + i\hbar\gamma^{\rm p}$.

The time evolution of the expectation values in a rotating frame with the drive frequency $\omega_{\rm d}$ reads~\cite{savasta1996quantum,katsch2019theory,suppmat}
\begin{align}
\label{eq:eom}
\begin{split}
  -i\hbar\partial_t \ev*{a_{\zeta,0}^\dagger} &= (\tilde{E}^{\rm p}_0-\hbar\omega_{\rm d})\ev*{a_{\zeta,0}^\dagger} + \Omega_0 \ev*{P^\dagger_{\zeta,0}}
  +i\hbar\sqrt{2\gamma^{\rm p}}\!\ev*{a^{\rm in\dagger}_{\zeta}}
  \\
  -i\hbar\partial_t\ev*{P_{\zeta,0}^\dagger} &= (\tilde{E}^{\rm x}_0-\hbar\omega_{\rm d})\ev*{P^{\dagger}_{\zeta,0}} + \Omega_0\ev*{a_{\zeta,0}^\dagger}
  \\ &-
  \sum_\bq \tilde{\Omega}_\bq\qty(\mathcal{C}^{\zeta\zeta'}_\bq + \delta_{\bq,0}\ev*{a_{\zeta,0}^\dagger}\ev*{P_{\zeta,0}^\dagger})\ev*{P_{\zeta,0}}
  \\ &+ W^0 \abs*{\ev*{P_{\zeta,0}^\dagger}}^2\ev*{P_{\zeta,0}^\dagger}
  +\sum_{\mu\zeta'\pm} W_{\mu}^{\pm}
  \mathcal{B}_{\mu,\pm}^{\zeta\zeta'} \ev*{P_{\zeta',0}}.
\\
-i\hbar\partial_t \mathcal{B}_{\mu,\pm}^{\zeta\zeta'} &= (\tilde{E}^{\rm xx}_{\mu,\pm}-2\hbar\omega_{\rm d})\mathcal{B}_{\mu,\pm}^{\zeta\zeta'}
+
\frac{1}{2}(1\pm\delta_{\zeta\zeta'})
\\ &\hspace{-0.5cm}\times
\{\widebar{W}^\pm_{\mu}
\ev*{P^\dagger_{\zeta,0}}\ev*{P^\dagger_{\zeta',0}}
+
\sum_{\bq}[
\widebar{\Omega}_{\mu,-\bq}^\pm
\mathcal{C}^{\zeta'\zeta}_{-\bq}
+
\widebar{\Omega}_{\mu,\bq}^\pm
\mathcal{C}^{\zeta\zeta'}_\bq
]
\}
\\
-i\hbar\partial_t \mathcal{C}^{\zeta\zeta'}_\bq &=
(\tilde{E}^{\rm p}_{\bq}+\tilde{E}^{\rm x}_{\bq}
 - 2\hbar\omega_{\rm d})\mathcal{C}^{\zeta\zeta'}_\bq
 \\
 &+ \Omega_{\bq} \mathcal{D}^{\zeta\zeta'}_{\bq}
 - \frac{1}{2}\delta_{\zeta\zeta'}\tilde{\Omega}_\bq \ev*{P^\dagger_{\zeta,0}}^2
 + \sum_{\mu\pm} \Omega_{\mu,\bq}^{\pm}\mathcal{B}_{\mu,\pm}^{\zeta\zeta'}
\\
-i\hbar\partial_t \mathcal{D}^{\zeta\zeta'}_\bq &=
2(\tilde{E}^{\rm p}_\bq - \hbar\omega_{\rm d})
\mathcal{D}^{\zeta\zeta'}_\bq
+ \Omega_\bq \mathcal{C}^{\zeta'\zeta}_{-\bq}
+ \Omega_{-\bq}\mathcal{C}^{\zeta\zeta'}_\bq
.
\end{split}
\end{align}
The first term in every equation describes free evolution, and the remaining terms describe couplings as illustrated in Fig.~\ref{fig:2}(c). For the photon amplitude $\ev*{a^\dagger_{\zeta,0}}$, the second term describes linear coupling to the exciton with rate $\Omega_0$ (where $2\Omega_0$ is the vacuum Rabi splitting) and the last term is input-field driving through the cavity mode, where the input-field expectation value is related to the input power $\mathcal{P}_{\rm in}$ and polarisation as~\cite{steck2007quantum} $\ev*{\mathbf{a}^{\rm in\dagger}}=\bm{\lambda}_{\rm in} [\mathcal{P}_{\rm in}/E^{\rm p}_0]^{1/2}$. For $\ev*{P^\dagger_{\zeta,0}}$, the second term describes linear coupling to photons. The third term stems from the fermionic substructure of excitons and generates nonlinear saturation of the light-matter interaction due to Pauli blocking $\tilde{\Omega}_\bq$. The last two terms describe uncorrelated mean-field Coulomb interactions ($W^0$) and beyond that Coulomb interactions with the biexcitonic correlations ($W^\pm_\mu$).

For the biexcitonic correlations $\mathcal{B}^{\zeta\zeta'}_{\mu,\pm}$, the second term contains Coulomb-scattering of uncorrelated excitons ($\widebar{W}^\pm_\mu$). \edit{For the bound biexciton ($\mu={\rm b}$), this corresponds to the process depicted in Fig.~\ref{fig:1}(c)}. The third term describes coupling to exciton-photon correlations through the light-matter interaction ($\widebar{\Omega}_{\mu,\bq}^\pm$).
For the exciton-photon correlations $\mathcal{C}^{\zeta\zeta'}_\bq$, the second term describes linear coupling to two-photon correlations by exchanging an exciton with a photon ($\Omega_0$). The third term describes nonlinear scattering of two uncorrelated excitons ($\tilde{\Omega}_\bq$), and the last term describes coupling to biexcitonic correlations via optical fields ($\Omega^\pm_{\mu,\bq}$). The second and third terms in the equation of motion for two-photon correlations $\mathcal{D}^{\zeta\zeta'}_\bq$ describe coupling to exciton-photon correlations by exchanging a photon with an exciton through the light-matter coupling $\Omega_0$. All definitions are given in the Supplementary Material~\cite{suppmat}.

As we will show, the bound biexciton $\mathcal{B}_{\mathrm{b},-}^{\zeta\zeta'}$ is the dominating contribution to the parametric gain as depicted in Fig.~\ref{fig:1}(d). This effect is absent in conventional third-order non-linear materials driven far off-resonantly and in two-level systems, which only have Pauli-blocking nonlinearity.

\begin{figure}[t]
\centering
\includegraphics[width=\columnwidth]{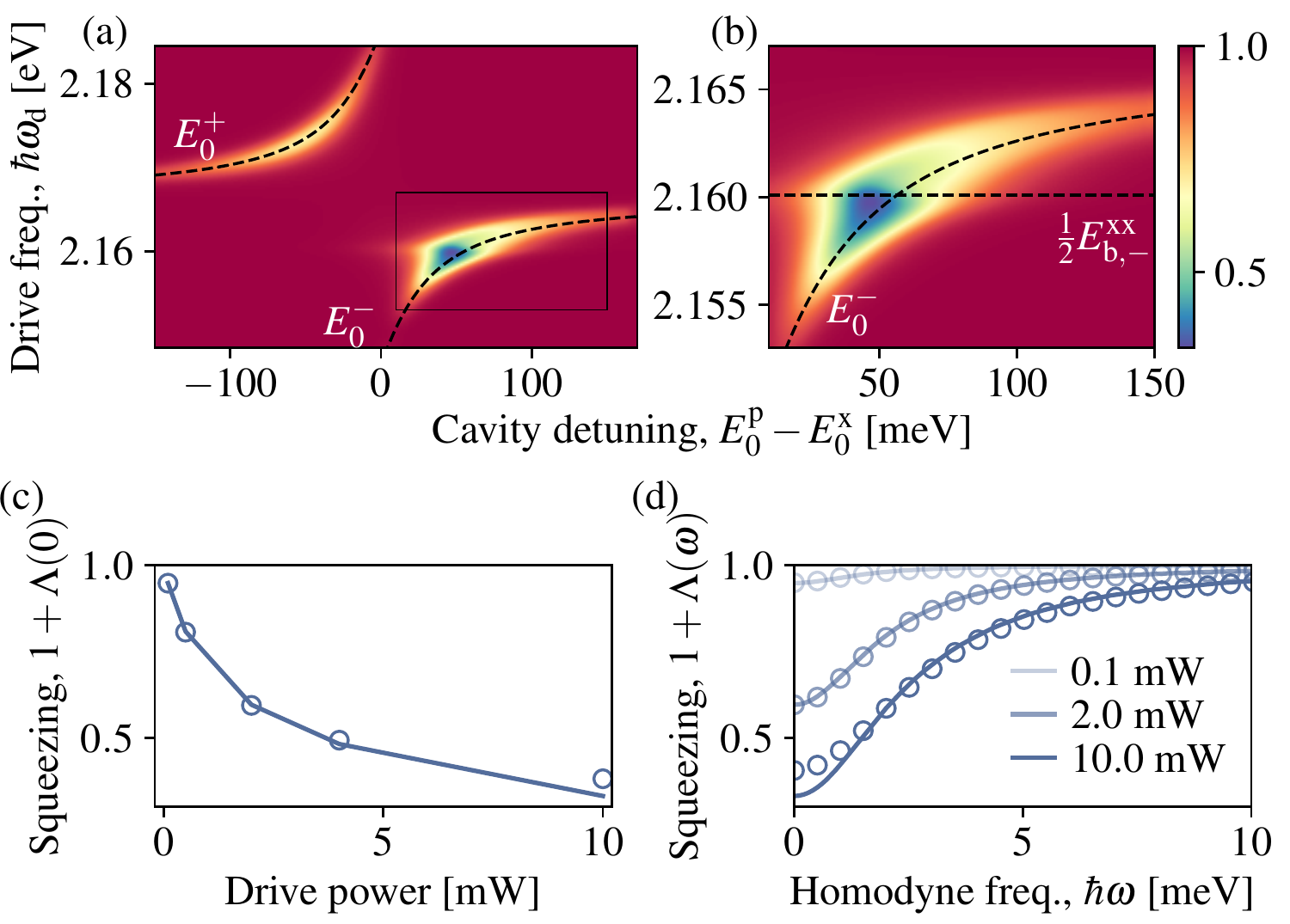}
  \caption{(a) Squeezing as $1+\Lambda(0)$ in the co-polarised output channel versus cavity-exciton detuning and drive frequency for a microcavity with hBN-encapsulated monolayer $\mathrm{MoS_2}$, at 10 mW driving power. Cavity parameters: $\Omega_0 = 20\mathrm{\:meV},\; \hbar\gamma^{\rm p} = 9\mathrm{\:meV}$. The temperature is $30\mathrm{\:K}$, leading to $\hbar\gamma^{\rm x}=0.8\mathrm{\:meV}$. The laser spot area is $9\mathrm{\:\mu m^2}$.
  (b) Zoom-in of region indicated by the rectangle in panel (a). (c) Squeezing at the numerically optimized cavity and driving frequecies versus driving power. Solid lines and open circles signify the co- and cross-polarized output channels.
  (d) Homodyne squeezing spectrum at the optimal driving frequency and cavity detuning.
  }
  \label{fig:3}
\end{figure}

To calculate the squeezing spectrum, we employ a Heisenberg-Langevin approach, where the time-dependent exciton and photon fluctuation operators $\delta P^\dagger_{\zeta,\bq}=P^\dagger_{\zeta,\bq}-\ev*{P^\dagger_{\zeta,\bq}}_{\rm s}$ and $\delta a^\dagger_{\zeta,\bq}=a^\dagger_{\zeta,\bq}-\ev*{a^\dagger_{\zeta,\bq}}_{\rm s}$ are defined with respect to the steady state expectation values of Eq.~\eqref{eq:eom}. Multiparticle fluctuations $\delta\mathcal{B}_{\mu,\pm}^{\zeta\zeta'},\;\delta\mathcal{C}^{\zeta\zeta'}_\bq$ and $\delta\mathcal{D}^{\zeta\zeta'}_\bq$ are defined similarly.

Due to the fluctuation-dissipation theorem~\cite{lax1966quantum,franke2019quantization,steck2007quantum,portolan2008nonequilibrium}, the Heisenberg-Langevin equations are driven by input noise operators for the photons ($\delta a^{\rm in}_\zeta$) and excitons ($\delta P_{\zeta,\bq}^{\rm in}$). Assuming that the fluctuations around their steady-state values are small, \edit{we approximate the equations of motion by their linearized form by removing products of fluctuation operators} and Fourier transform to obtain
\begin{align}
\label{eq:fluctuation-equations}
\begin{split}
  -&(\hbar\omega-\hbar\omega_{\rm d}+\tilde{E}^{\rm p}_0)\delta \mathbf{a}_0^\dagger(\omega)
  = \Omega_0\delta\mathbf{P}^\dagger_0(\omega) + i\hbar\sqrt{2\gamma^{\rm p}}\delta\mathbf{a}^{\rm in\dagger}(\omega)
\\
-&[\hbar\omega-\hbar\omega_{\rm d}+\tilde{E}^{\rm x}_0 + \bm{\Sigma}(\omega)]\delta\mathbf{P}^\dagger_0(\omega)
=  \bm{\Omega}_0^{\rm r}(\omega)\delta\mathbf{a}^\dagger_0(\omega) \\ &\hspace{3cm}+ \bm{\Delta}\:\delta \mathbf{P}_0(\omega)
 + i\hbar\sqrt{2\bm{\Gamma}^{\rm x}(\omega)}\delta\mathbf{P}^{\rm in\dagger}_0(\omega)
\\ &\hspace{4cm}+ i\hbar\sqrt{2\bm{\Gamma}^{\rm p}(\omega)}\delta\mathbf{a}^{\rm in\dagger}(\omega),
\end{split}
\end{align}
where bold symbols denote vectors and matrices in the $\zeta$-basis. The multiparticle fluctuation equations have been formally solved, leading to the self energy $\bm{\Sigma}$, the renormalised input field couplings $\bm{\Gamma}^{\rm x/p}$ and the renormalised coupling $\bm{\Omega}^{\rm r}_0$ (see Supplementary Material~\cite{suppmat}). Eqs.~\eqref{eq:fluctuation-equations} are solved in order to calculate the squeezing spectrum $\Lambda(\omega)$. Importantly, $\bm{\Delta}$ in Eq.~\eqref{eq:fluctuation-equations} is the parametric gain that generates squeezing, which arises from the nonlinear response and takes the form
\begin{align}
\label{eq:Delta-matrix}
  \begin{split}
    \Delta_{\zeta\zeta'} = \delta_{\zeta,\zeta'}W^0 &\ev*{P_{\zeta,0}^\dagger}^2
    + \sum_{\mu\pm} W_{\mu}^\pm \mathcal{B}^{\zeta\zeta'}_{\mu\pm}
    \\
    &-\delta_{\zeta,\zeta'}\sum_\bq \tilde{\Omega}_\bq\big[ \mathcal{C}^{\zeta\zeta'}_\bq
    +
    \delta_{\bq,0}\ev*{a_{\zeta,0}^\dagger}\ev*{P_{\zeta,0}^\dagger} \big].
  \end{split}
\end{align}
This quantity, which in Eq.~\eqref{eq:fluctuation-equations} couples $\delta P^\dagger_{\zeta,0}$ to the conjugate field $\delta P_{\zeta',0}$, is analogous to the two-photon pump rate in the well-known degenerate parametric amplifier~\cite{gardiner2004quantum}. The three terms contributing to the parametric gain in Eq.~\eqref{eq:Delta-matrix} are generated by mean-field exciton Coulomb interaction, biexcitonic correlations and Pauli-blocking, respectively.

\begin{figure}
\includegraphics[width=\columnwidth]{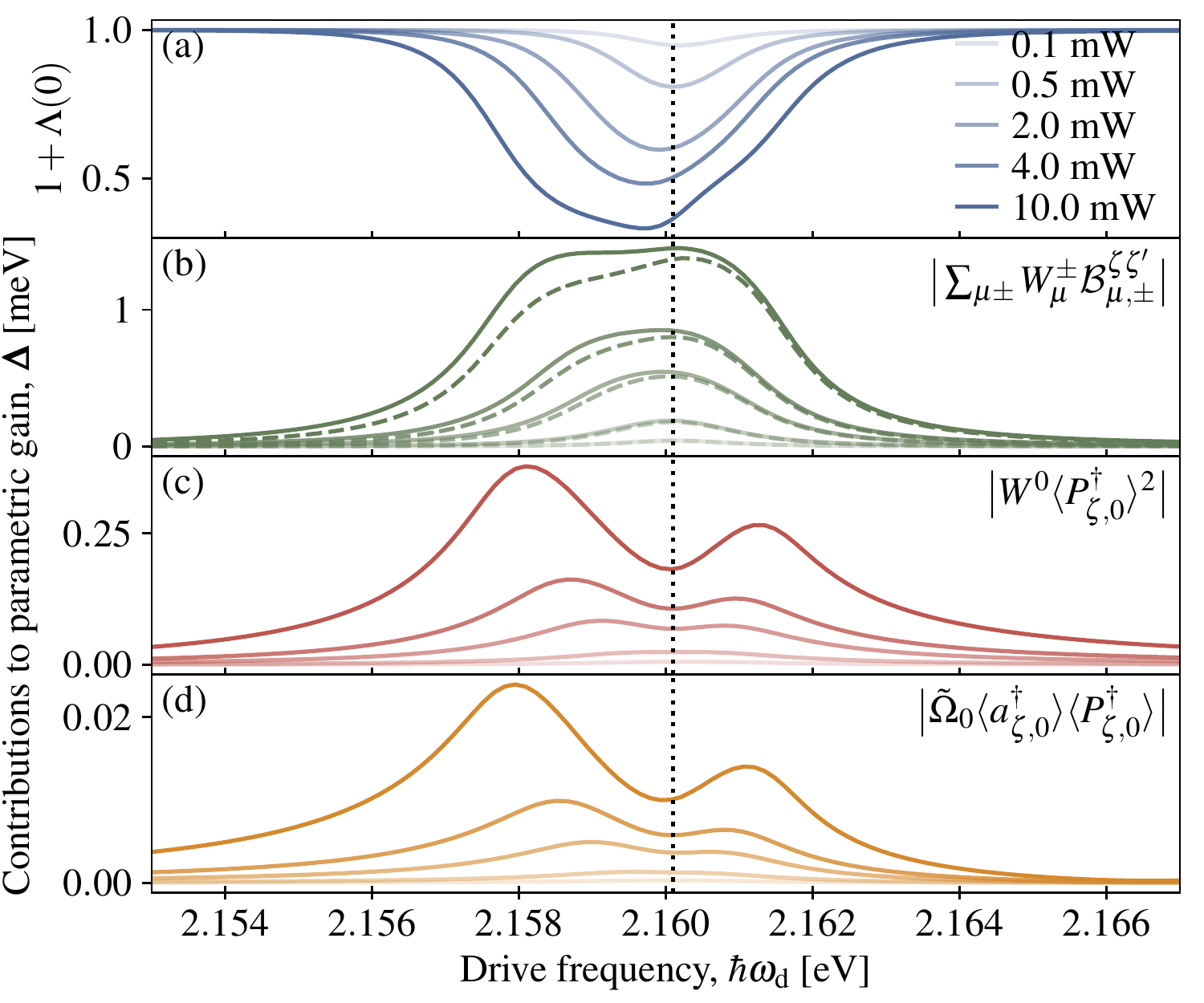}
  \caption{(a) Co-polarised squeezing versus driving frequency for optimized cavity detuning at different driving power levels. (b)-(d) Contributions to parametric gain from biexcitonic correlations (b), mean-field Coulomb interaction (c) and Pauli blocking (d). The contributions in (c) and (d) only have diagonal contributions in the $\zeta$-basis, which are equal for linearly polarized driving. In (b) the diagonal and off-diagonal contributions have been added. The dashed lines in (b) show the contibution from the bound biexciton alone ($\mu=\mathrm{b}$, strictly off-diagonal). The vertical dotted line indicates the two-photon resonance $\frac{1}{2}E^{\rm xx}_{\rm b,-}$.}
  \label{fig:4}
\end{figure}

{\it Results.\textemdash}
Fig.~\ref{fig:3}(a)--(b) shows the squeezing as the quadrature noise at zero homodyne detection frequency $1+\Lambda(0)$ in the co-polarized output as a function of the drive frequency and the exciton-cavity detuning at 10 mW driving power, for hBN-encapsulated monolayer ${\rm MoS_2}$, cavity parameters compatible with fabricated devices~\cite{liu2015strong,anton2021bosonic,suppmat} and a temperature of 30 K. \edit{The phonon-induced exciton dephasing $\gamma^{\rm x}$ should be significantly smaller than the photon outcoupling rate $\gamma^{\rm p}$, such that polaritons are coupled out of the cavity before they scatter with phonons, making cryogenic temperatures necessary.}

A value of $1+\Lambda(0)=1$ corresponds to the shot-noise level, i.e. no squeezing, whereas $1+\Lambda(0)=0$ corresponds to complete elimination of noise in one quadrature, i.e. perfect squeezing.

The dominating response is around the polariton energies $E^\pm_0 = \frac{1}{2}\{E^{\rm p}_0+E^{\rm x}_0 \pm [(E^{\rm p}_0-E^{\rm x}_0)^2 + 4\Omega_0^2]^{1/2}\}$, and a particularly strong squeezing is seen where the lower polariton branch is two-photon resonant with the bound biexciton, $E^-_0\simeq\frac{1}{2}E^{{\rm xx}}_{\rm b,-}$.
The dependence of squeezing on the driving power is shown in Fig.~\ref{fig:3}(c) at the optimal cavity and driving frequencies. Fig.~\ref{fig:3}(d) shows the squeezing as a function of homodyne detection frequency, demonstrating a bandwidth of several meV. This large bandwidth stems from the cavity outcoupling rate~\cite{hoff2015integrated} $\gamma^{\rm p}$ and exciton dephasing $\gamma^{\rm x}$ which are also in the meV range~\cite{selig2016excitonic}.

The input power of 1-10 mW is an order of magnitude below the typical range of 50-100 mW required to generate comparable squeezing levels in state-of-the art on-chip devices with conventional third-order nonlinear media~\cite{zhao2020near,zhang2021squeezed}. Specifically, in Ref.~\cite{seifoory2022degenerate}, squeezing in an optimized $\mathrm{Si_3N_4}$ microring resonator is predicted down to 84\% ($-0.75\:\mathrm{dB}$) for a driving power of 10 mW, whereas we predict 33\% ($-4.8\:\mathrm{dB}$) for the same power.

To understand the dominating physical processes responsible for squeezing, we show in Fig.~\ref{fig:4} the squeezing as a function of drive frequency along with the three contributions to the parametric gain $\bm{\Delta}$ from Eq.~\eqref{eq:Delta-matrix}. The cavity detuning has been chosen by numerically optimizing the squeezing as in Fig.~\ref{fig:3}. The contribution from exciton-photon correlations $\mathcal{C}_\bq^{\zeta\zeta'}$ was found to be negligible and is not shown here. The contributions from biexcitonic correlations exceed the mean-field Coulomb and Pauli-blocking by almost an order of magnitude. We can single out the contribution from the bound biexciton ($\mu=\mathrm{b}$) in Eq.~\eqref{eq:Delta-matrix}, [dashed lines in Fig.~\ref{fig:4}(b)], which accounts for more than 80\% of the total parametric gain in the frequency region with strongest squeezing. \edit{Thus, resonant Coulomb-mediated biexciton formation as shown in Fig.~\ref{fig:1}(c)-(d) is the main contribution to the parametric gain and squeezing.}

\emph{Conclusion.\textemdash}In conclusion, we have presented a theoretical analysis of the generation of quadrature-squeezed light using the biexcitonic resonance in an atomically thin semiconductor coupled to an optical microcavity. We have shown that siginificant levels of broadband squeezing can be generated with very low input power levels of the order of 1-10 mW.

\edit{A previous experimental investigation~\cite{shimano2002efficient} measured parametric gain from biexcitons in a bulk CuCl microcavity in the near-UV spectral range. The squeezing level at pump power equivalent to 63 mW for the spot size considered here was inferred to 0.63\% ($-2\:\mathrm{dB}$), although not directly measured. Furthermore, we note that ZnO quantum wells with biexciton binding energies around 15 meV~\cite{ko2000biexciton} are another interesting platform to potentially observe the predicted squeezing mechanism in the near-UV spectrum.}

An interesting extension of the use of atomically thin semiconductors for quadrature squeezing is to  introduce an electromagnetic nanoresonator with tight in-plane optical confinement~\cite{wen2017room,zheng2017manipulating,kleemann2017strong, cuadra2018observation, stuhrenberg2018strong, han2018rabi, geisler2019single, qin2020revealing}. Such structures could potentially enhance the efficiency, because the in-plane confinement leads to a stronger nonlinear response~\cite{denning2022quantum,denning2022cavity}.

\begin{acknowledgments}
We thank Erik Bærentsen and Malte Selig for insightful discussions.
E.V.D. acknowledges support from Independent Research Fund Denmark through an International Postdoc Fellowship (Grant No. 0164-00014B). F.K. and A.K. gratefully acknowledge support from the Deutsche Forschungsgemeinschaft through Projects No.~420760124 (\mbox{KN 427/11-1}) and No. 163436311---SFB~910 (Project B1).
\end{acknowledgments}

\pagebreak
\widetext
\begin{center}
\textbf{\large Supplemental Materials: Efficient quadrature-squeezing from biexcitonic parametric gain in atomically thin semiconductors}
\end{center}
\setcounter{equation}{0}
\setcounter{section}{0}
\setcounter{figure}{0}
\setcounter{table}{0}
\setcounter{page}{1}
\makeatletter
\renewcommand{\theequation}{S\arabic{equation}}
\renewcommand*{\thesection}{S~\Roman{section}}
\renewcommand{\thefigure}{S\arabic{figure}}
\renewcommand{\bibnumfmt}[1]{[S#1]}
\renewcommand{\citenumfont}[1]{S#1}

\section{Hamiltonian}
The noninteracting Hamiltonian $H_0=H_{\rm e}+H_{\rm p}+H_{\rm e-p}$ contains contributions from free electrons in the valence and conduction bands ($H_{\rm e}$), free photons ($H_{\rm p}$) and electron-photon coupling ($H_{\rm e-p}$). The electron part is given by
\begin{align}
  H_{\rm e} = \sum_{\zeta\bk} \qty(E_\bk^{\rm c} c_{\zeta,\bk}^\dagger c_{\zeta,\bk}
  + E_\bk^{\rm v} v_{\zeta,\bk}^\dagger v_{\zeta,\bk}),
\end{align}
where $E^{\rm c}_\bk = E^{\rm c}_0 + \hbar^2 k^2/2m_{\rm e}$ and $E^{\rm v}_\bk = E^{\rm v}_0 - \hbar^2 k^2/2m_{\rm h}$ are the band energies of the conduction and valence bands with $E^{\rm c}_0-E^{\rm v}_0$ the quasiparticle bandgap and $m_{\rm e}$ ($m_{\rm h}$) the effective electron (hole) mass. Since the considered band energies are equal for the $K$ and $K'$ valleys, there is no $\zeta$-index on $E^{\rm c}_\bk$ and $E^{\rm v}_\bk$.

The photon part is given by
\begin{align}
  H_{\rm p} = \sum_{\zeta\bk} E_\bk^{\rm p} a_{\zeta\bk}^\dagger a_{\zeta\bk},
\end{align}
where the cavity photon energy is given by~\cite{skolnick1998strongSI,osgood2021dielectricSI} $E^{\rm p}_\bk = \hbar[\omega_{\rm p,0}^2 + (c k/\widebar{n})^2]^{1/2}$, where $\omega_{\rm p,0}$ is the resonance frequency of the cavity mode, $c$ is the vacuum speed of light and $\widebar{n}$ is the effective cavity refractive index.

Within the rotating-wave approximation, the electron-photon coupling is given by
\begin{align}
\label{eq:H-e-p}
  H_{\rm e-p} = \sum_{\zeta\bk\bq} \qty(A_\bq c_{\zeta,\bk+\bq}^\dagger v_{\zeta,\bk} a_{\zeta,\bq}
  + A_\bq^* a_{\zeta,\bq}^\dagger v_{\zeta,\bk}^\dagger c_{\zeta,\bk+\bq} ),
\end{align}
where $A_\bq=\sqrt{\omega_0/\omega_\bq}A_0$ is the light-matter coupling coefficient, where $A_0$ depends on the out-of-plane confinement of the cavity mode and the valence-conduction band Bloch matrix element of the atomically thin semiconductor~\cite{kira1999quantumSI,denning2022quantumSI}. \edit{The rotating-wave approximation is valid, when the exciton-photon coupling strength is significantly smaller than the photon and exciton energies. In typical microcavities with atomically-thin transition-metal dichalcogenides, the coupling strength is on the order of 10 meV~\cite{liu2015strongSI,dufferwiel2015excitonSI}, whereas the exciton and cavity photon energies are around 2 eV; thus the rotating wave approximation is justified.}

The Coulomb Hamiltonian is given by~\cite{katsch2018theorySI}
\begin{align}
\begin{split}
  H_{\rm C} = \frac{1}{2}\sum_{\bk_1\bk_2\bq}  \sum_{\zeta_1\zeta_2} V_\bq\Big(&c^\dagger_{\zeta_1,\bk_1+\bq}c^\dagger_{\zeta_2,\bk_2-\bq}
  c_{\zeta_2,\bk_2}c_{\zeta_1,\bk_1}
  + v^\dagger_{\zeta_1,\bk_1+\bq}v^\dagger_{\zeta_2,\bk_2-\bq}
  v_{\zeta_2,\bk_2}v_{\zeta_1,\bk_1}
\\  &+ c^\dagger_{\zeta_1,\bk_1+\bq}v^\dagger_{\zeta_2,\bk_2-\bq}
  v_{\zeta_2,\bk_2}c_{\zeta_1,\bk_1}
  + v^\dagger_{\zeta_1,\bk_1+\bq}c^\dagger_{\zeta_2,\bk_2-\bq}
  c_{\zeta_2,\bk_2}v_{\zeta_1,\bk_1}\Big).
\end{split}
\end{align}
Here, $V_\bq= e_0^2[2S\epsilon_0\epsilon_\bq q]^{-1}$ is the screened 2D Coulomb potential, where $e_0$ is the elementary charge, $S$ is the quantization surface area, $\epsilon_0$ is the vacuum permittivity, and $\epsilon_\bq$ is the dielectric function which is described in Sec.~\ref{sec:methods}.
Here, we have neglected inter- and intravalley exchange interactions, which have previously been shown to be significantly weaker than the direct interaction in monolayer transition-metal dichalcogenides~\cite{katsch2018theorySI, katsch2019theorySI}. We note that such exchange effects can give rise to biexciton fine structure~\cite{steinhoff2020dynamicalSI} and corrections to the biexciton binding energy~\cite{kwong2021effectSI}.

In the end, all coupling coefficients and scattering matrices in the equations of motion are independent of the quantization surface area $S$, which cancels out in the final expressions when amplitudes $\ev*{a^\dagger}$ and $\ev*{c^\dagger v}$ are expressed in surface-density units $\ev*{a^\dagger}/\sqrt{S}$ and $\ev*{c^\dagger v}/\sqrt{S}$. Only the input-field driving term contains explicit reference to $S$, when converting the driving power $\mathcal{P}_{\rm in}$ to surface-density units $\mathcal{P}_{\rm in}\rightarrow\mathcal{P}_{\rm in}/S$. For the numerical calculations, $S$ is then taken as the laser spot area.

\section{Equations of motion for expectation values}
\label{sec:dct-eom}
For the time evolution of the relevant expectation values, we use the Heisenberg equation of motion $-i\hbar\partial_t Q = [H, Q]$. For the photon coherence we then have
\begin{align}
-i\hbar\partial_t &\ev*{a_{\zeta,0}^\dagger} =
E^{\rm p}_0 \ev*{a^\dagger_{\zeta,0}} + \sum_{\bk} A_0 \ev*{c_{\zeta,\bk}^\dagger v_{\zeta,\bk}}
+ i\hbar\sqrt{2\gamma^{\rm p}} \ev*{a^{\rm in\dagger}_{\zeta}},
\end{align}
where the input-field term is derived from input-output theory~\cite{gardiner2004quantumSI,steck2007quantumSI}. For the semiconductor polarization, we have
\begin{align}
\label{eq:pair-eom}
\begin{split}
  -i\hbar\partial_t & \ev*{c_{\zeta\bk}^\dagger v_{\zeta\bk}}
  = A_0^* \ev*{a_{\zeta,0}^\dagger}
- \sum_{\bq}
  \Big(
  A_{\zeta,\bq}^{*}
  \ev*{a_{\zeta,\bq}^\dagger c^\dagger_{\zeta\bk} v_{\zeta,\bk+\bq}}
   \ev*{v^\dagger_{\zeta,\bk+\bq} c_{\zeta\bk+\bq}}
  + A_{-\bq}^{*} \ev*{a_{\zeta,-\bq}^\dagger
   c^\dagger_{\zeta\bk+\bq} v_{\zeta\bk}}
   \ev*{v^\dagger_{\zeta\bk+\bq}c_{\zeta\bk+\bq}}\Big)
  \\ &+
  (E_\bk^{\rm c}-E^{\rm v}_\bk)\ev*{c^\dagger_{\zeta\bk}v_{\zeta\bk}} - \sum_{\bq}V_\bq \ev*{c_{\zeta\bk+\bq}^\dagger v_{\zeta\bk+\bq}}
  \\
  &+ \sum_{\zeta'\bk'\bq} V_\bq
    \qty[\ev*{c^\dagger_{\zeta\bk+\bq}v_{\zeta\bk}c_{\zeta'\bk'}^\dagger
    v_{\zeta'\bk'+\bq}} +
    \ev*{c^\dagger_{\zeta\bk}v_{\zeta\bk+\bq}c_{\zeta'\bk'+\bq}^\dagger v_{\zeta'\bk'}}]
    \qty[\ev*{v^\dagger_{\zeta'\bk'+\bq}c_{\zeta'\bk'+\bq}} - \ev*{v^\dagger_{\zeta'\bk'}c_{\zeta'\bk'}}]
  \end{split}
\end{align}
Note that electron-hole coherences with different momenta $\ev*{c^\dagger_{\zeta\bk}v_{\zeta\bk'}},\; \bk\neq\bk'$, are identically zero due to the normal incidence of the driving field.
To derive this equation, we have first expanded electron and hole densities in terms of pair operators using a unit-operator expansion method~\cite{ivanov1993selfSI,katsch2018theorySI}
\begin{align}
\label{eq:unit-operator-expansion}
\begin{split}
  c_{\zeta\bk}^\dagger c_{\zeta'\bk'} &= \sum_{\zeta_1\bk_1}c_{\zeta\bk}^\dagger v_{\zeta_1\bk_1}v_{\zeta_1\bk_1}^\dagger c_{\zeta'\bk'} -
   \frac{1}{2}\sum_{\zeta_1\bk_1}\sum_{\zeta_2\bk_2}\sum_{\zeta_3\bk_3}
  c_{\zeta\bk}^\dagger v_{\zeta_1\bk_1}
  c_{\zeta_2\bk_2}^\dagger v_{\zeta_3\bk_3}
  v_{\zeta_3\bk_3}^\dagger c_{\zeta_2\bk_2}
  v_{\zeta_1\bk_1}^\dagger c_{\zeta'\bk'}
  + \cdots
  \\
  v_{\zeta\bk} v_{\zeta'\bk'}^\dagger &= \sum_{\zeta_1\bk_1}c_{\zeta_1\bk_1}^\dagger v_{\zeta\bk}v_{\zeta'\bk'}^\dagger c_{\zeta_1\bk_1} -
   \frac{1}{2}\sum_{\zeta_1\bk_1}\sum_{\zeta_2\bk_2}\sum_{\zeta_3\bk_3}
  c_{\zeta_1\bk_1}^\dagger v_{\zeta\bk}
  c_{\zeta_2\bk_2}^\dagger v_{\zeta_3\bk_3}
  v_{\zeta_3\bk_3}^\dagger c_{\zeta_2\bk_2}
  v_{\zeta'\bk'}^\dagger c_{\zeta_1\bk_1}
  + \cdots
\end{split}
\end{align}
which is valid when the only source of electrons and holes is optical excitation, \edit{i.e. when no doping or electrical injection is present.}
We then used the dynamics-controlled truncation (DCT) scheme~\cite{axt1994dynamicsSI,lindberg1994chiSI} to perturbatively expand the equations of motion to third order in the driving field $a^{\rm in}_\zeta$, meaning that only terms with up to three normal-ordered pair or photon operators are kept.

Furthermore, the third-order terms have been factorized as $\ev*{c^\dagger v c^\dagger v v^\dagger c}=\ev*{c^\dagger v c^\dagger v}\ev*{v^\dagger c}$ and $\ev*{a^\dagger c^\dagger v v^\dagger c}=\ev*{a^\dagger c^\dagger v}\ev*{v^\dagger c}$. \edit{This factorization is valid for third-order DCT in the coherent regime~\cite{savasta1996quantumSI}, i.e. when the only source of electrons and holes is excitation with coherent light near the resonances of the system, and when incoherent scattering processes e.g. via phonons can be neglected. Phonon scattering is later included phenomenologically through a dephasing rate, which is obtained from a separate self-consistent microscopic calculation~\cite{selig2016excitonicSI}. This means that the validity of our approach is limited to the regime where cavity outcoupling ($\gamma^{\rm p}$) dominates over exciton dephasing ($\gamma^{\rm x}$), which guarantees that polaritons will be outcoupled before significant dephasing takes place.}

At this point, we introduce the exciton wavefunction $\phi_{\bk}^n$ as the solution to the Wannier equation~\cite{wannier1937structureSI,sham1966manySI,kira2006manySI}
\begin{align}
\label{eq:wannier}
  \qty(E^{\rm c}_{0}-E^{\rm v}_0 + \frac{\hbar^2 k^2}{2m})\phi^n_\bk
  - \sum_\bq V_\bq \phi^n_{\bk+\bq} = E^{\rm x}_0 \phi^n_\bk,
\end{align}
where $m = [m_{\rm e}^{-1} + m_{\rm h}^{-1}]^{-1}$ is the reduced mass.  The wavefunctions are orthonormal, such that $\sum_{n} \phi^{n*}_{\bk}\phi^{n}_{\bk'} = \delta_{\bk\bk'}$ and $\sum_{\bk}\phi^{n*}_\bk\phi^{n'}_\bk=\delta_{nn'}$. Using the exciton wavefunction set, we can express electron-hole pair operators in terms of exciton operators, $P_{\zeta\zeta',\bq}^n$ through the relations
\begin{align}
\label{eq:exciton-expansion}
\begin{split}
  P_{\zeta\zeta',\bq}^{n\dagger} &= \sum_\bk \phi^{n}_{\bk}
  c^\dagger_{\zeta,\bk+\alpha\bq} v_{\zeta',\bk-\beta\bq}
  \\
  c^\dagger_{\zeta,\bk}v_{\zeta',\bk'} &= \sum_n \phi^{n*}_{\beta\bk + \alpha\bk'} P^{n^\dagger}_{\zeta\zeta',\bk-\bk'},
\end{split}
\end{align}
where $\bq$ is the exciton center-of-mass momentum and $\alpha=m_{\rm e}/(m_{\rm e}+m_{\rm h}),\;\beta=m_{\rm h}/(m_{\rm e}+m_{\rm h})$. In principle, the index $n$ runs over all solutions to Eq~\eqref{eq:wannier}. However, since the lowest-energy exciton ($n=\mathrm{1s}$) is separated from the next excitonic state by an energy gap of hundreds of meV, we truncate all summations over $n$ to only include $n=\mathrm{1s}$, thereby projecting the electron-hole pair space onto the 1s excitonic state. Thus, we shall omit the $n$-index on the exciton wavefunctions and operators. Furthermore, we will use the shorthand notation $P_{\zeta,\bq}:=P_{\zeta\zeta,\bq}$ for intravalley excitons ($\zeta'=\zeta$).

Next, the two-pair expectation values in Eq.~\eqref{eq:pair-eom} are separated into factorized and correlated parts, defined as
\begin{align}
\label{eq:two-pair-correlation}
  \ev*{c^\dagger_{\zeta_1,\bk}v_{\zeta_1,\bk+\bq}c^\dagger_{\zeta_2,\bk'+\bq}v_{\zeta_2,\bk'}}^{\rm c} := \ev*{c^\dagger_{\zeta_1,\bk}v_{\zeta_1,\bk+\bq}c^\dagger_{\zeta_2,\bk'+\bq}v_{\zeta_2,\bk'}}
  -\ev*{c^\dagger_{\zeta_1,\bk}v_{\zeta_1,\bk+\bq}}\ev*{c^\dagger_{\zeta_2,\bk'+\bq}v_{\zeta_2,\bk'}}
  +\ev*{c^\dagger_{\zeta_2,\bk'+\bq}v_{\zeta_1,\bk+\bq}}\ev*{c^\dagger_{\zeta_1,\bk}v_{\zeta_2,\bk'}}.
\end{align}
These correlated two-pair expectation values are then projected on the 1s-exciton subspace as
\begin{align}
  \ev*{c^\dagger_{\zeta_1,\bk}v_{\zeta_1,\bk+\bq}c_{\zeta_2,\bk'+\bq}^\dagger v_{\zeta_2,\bk'}}^{\rm c} = \frac{1}{2}&\Big[\phi^*_{\bk + \alpha\bq}\phi^*_{\bk'+\beta\bq}
  \ev*{P^{\dagger}_{\zeta_1\zeta_1,-\bq} P^{\dagger}_{\zeta_2\zeta_2,\bq}}^{\rm c}
    - \phi^*_{\beta\bk'+ \alpha\bk +\bq}\phi^*_{\beta\bk + \alpha\bk'}
  \ev*{P^{\dagger}_{\zeta_2\zeta_1,\bk'-\bk} P^{\dagger}_{\zeta_1\zeta_2,\bk-\bk'}}^{\rm c}\Big],
\end{align}
where the correlated two-exciton expectation values $\ev*{P_1^\dagger P_2^\dagger}^{\rm c}$ are defined to obey Eqs.~\eqref{eq:two-pair-correlation} and~\eqref{eq:exciton-expansion}. We then define the singlet ($-$) and triplet ($+$) correlations as $\tilde{\mathcal{B}}_{\bq,\pm}^{\zeta\zeta'} = \frac{1}{4}(\ev*{P_{\zeta\zeta,\bq}^\dagger P_{\zeta'\zeta',-\bq}^\dagger}^{\rm c}\pm\ev*{P_{\zeta'\zeta,\bq}^\dagger P_{\zeta\zeta',-\bq}^\dagger}^{\rm c})$, such that
\begin{align}
\label{eq:B-tilde-expansion}
\ev*{c^\dagger_{\zeta_1,\bk}v_{\zeta_1,\bk+\bq}c_{\zeta_2,\bk'+\bq}^\dagger v_{\zeta_2,\bk'}}^{\rm c} = \sum_{\pm}\Big[\phi^*_{\bk + \alpha\bq}\phi^*_{\bk'+\beta\bq}
\tilde{\mathcal{B}}^{\zeta\zeta'}_{-\bq,\pm}
  \mp \phi^*_{\beta\bk'+ \alpha\bk +\bq}\phi^*_{\beta\bk + \alpha\bk'}
\tilde{\mathcal{B}}^{\zeta\zeta'}_{\bk'-\bk,\pm}\Big].
\end{align}
We note that this expansion is only possible when the effective masses of the involved holes (or electrons) are equal~\cite{katsch2019theorySI}. For the present case, this is not a limitation or even an approximation, because only the lowest-energy excitons are excited, whereby only the highest valence band and lowest conduction band are involved. Thus, no combinations of bands with unequal electron or hole masses occur.

Similarly, the correlated part of the electron-hole-photon expectation values is defined as $\ev*{a^\dagger_{\zeta,\bq}c^\dagger_{\zeta',\bk-\bq}v_{\zeta',\bk}}^{\rm c} = \ev*{a^\dagger_{\zeta,\bq}c^\dagger_{\zeta',\bk-\bq}v_{\zeta',\bk}} - \ev*{a^\dagger_{\zeta,\bq}}\ev*{c^\dagger_{\zeta',\bk-\bq}v_{\zeta',\bk}}$ and projected onto the 1s-exciton subspace as
$
  \mathcal{C}^{\zeta\zeta'}_\bq = \sum_\bk \phi_\bk
  \ev*{a^\dagger_{\zeta,\bq} c^\dagger_{\zeta',\bk-\alpha\bq} v_{\zeta',\bk+\beta\bq}}^{\rm c}.
$

We then project Eq.~\eqref{eq:pair-eom} onto the 1s-exciton subspace by multiplying by $\phi_\bk$ and summing over $\bk$. The resulting equation of motion for the excitonic amplitude then reads
\begin{align}
\label{eq:exciton-eom}
  -i\hbar\partial_t \ev*{P_{\zeta,0}^\dagger} &= E^{\rm x}_0  \ev*{P_{\zeta,0}^\dagger} + \Omega_0^*\ev*{a_{\zeta,0}^\dagger}
  - \sum_\bq \tilde{\Omega}_\bq^* \qty(\mathcal{C}^{\zeta\zeta}_\bq + \delta_{\bq,0}\ev*{a_{\zeta,0}^\dagger}\ev*{P_{\zeta,0}^\dagger})\ev*{P_{\zeta,0}}
+ W^{0*}\ev*{P_{\zeta,0}^\dagger}^2\ev*{P_{\zeta,0}}
  + \sum_{\zeta'\bq\pm} \tilde{W}_{\bq}^{\pm*} \tilde{\mathcal{B}}_{\bq,\pm}^{\zeta\zeta'}\ev*{P_{\zeta',0}},
\end{align}
where $\Omega_0 = A_0\sum_\bk\phi_\bk$ is the exciton-photon coupling strength and
\begin{align}
\label{eq:exciton-eom-coefficients}
\begin{split}
  \tilde{\Omega}_\bq &= \sum_{\bk_1} A_\bq(
  \phi_{\bk_1}^*
  \phi_{\bk_1+\alpha\bq}
  \phi_{\bk_1 + \bq}
  +
  \phi_{\bk_1+\bq}^*
  \phi_{\bk_1+\alpha\bq}
  \phi_{\bk_1}
  \Big)
  \\
  W^0 &= \sum_{\bk_1\bk_2} V_{\bk_2-\bk_1}
  \phi_{\bk_1}\phi_{\bk_1}
  (\phi_{\bk_1}^*-\phi_{\bk_2}^*)
  (\phi_{\bk_1}^*-\phi_{\bk_2}^*)
  \\
  \tilde{W}_\bq^\pm &=
  V_\bq\sum_{\bk_1\bk_2}
  \phi_{\bk_1}
  \phi_{\bk_2}
  (\phi^*_{\bk_1-\beta\bq} - \phi^*_{\nu_1,\bk_1+\alpha\bq})
  (\phi^*_{\bk_2+\beta\bq} - \phi^*_{\bk_2-\alpha\bq})
  \pm \sum_{\bk_1\bk_2}V_{\bk_1-\bk_2+(\alpha-\beta)\bq}
  \phi_{\bk_1}
  \phi_{\bk_2}
  (\phi^*_{\bk_1-\beta\bq} - \phi^*_{\bk_2-\alpha\bq})
  (\phi^*_{\bk_1+\alpha\bq} - \phi^*_{\bk_2+\beta\bq})
\end{split}
\end{align}
are the Pauli-blocking strength ($\tilde{\Omega}_\bq$), and the factorized ($W^0$) and correlated ($\tilde{W}_\bq^\pm$) exciton Coulomb interaction.

To proceed, we derive the equation of motion for $\mathcal{C}^{\zeta\zeta'}_\bq$. This is done by first calculating the equation of motion for the correlated electron-hole-photon amplitude $\ev*{a^\dagger_{\zeta,\bq}c_{\zeta',\bk-\bq}^\dagger v_{\zeta',\bk}}^{\rm c}$
\begin{align}
\label{eq:exciton-photon-amplitude-eom-1}
  \begin{split}
    -i\hbar\partial_t \ev*{a^\dagger_{\zeta,\bq} c_{\zeta'\bk-\bq}^\dagger v_{\zeta'\bk}}^{\rm c} &=
    \hbar\omega_\bq\ev*{a_{\zeta,\bq}^\dagger c_{\zeta',\bk-\bq}^\dagger v_{\zeta'\bk}}^{\rm c}
    +A^*_{-\bq} \mathcal{D}^{\zeta\zeta'}_{\bq}
    +
    \sum_{\bk'} A_{\bq}
    \qty[\ev*{c_{\zeta,\bk'+\bq}^\dagger v_{\zeta,\bk'}
    c_{\zeta',\bk-\bq}^\dagger v_{\zeta'\bk}} -
    \delta_{\bq 0}\ev*{c_{\zeta,\bk'}^\dagger v_{\zeta,\bk'}}
    \ev*{c_{\zeta',\bk}^\dagger v_{\zeta',\bk}}]
    \\ &+
    (E^{\rm c}_{\bk-\bq}-E^{\rm v}_\bk)
    \ev*{a_{\zeta,\bq}^\dagger c_{\zeta',\bk-\bq}^\dagger v_{\zeta',\bk}}^{\rm c}
    - \sum_\bp V_\bp \ev*{a_{\zeta,\bq}^\dagger c_{\zeta',\bk+\bp-\bq}^\dagger v_{\zeta',\bk+\bp}}^{\rm c},
  \end{split}
\end{align}
where $\mathcal{D}^{\zeta\zeta'}_\bq=\ev*{a^\dagger_{\zeta,\bq}a^\dagger_{\zeta',-\bq}}^{\rm c}:=\ev*{a^\dagger_{\zeta,\bq}a^\dagger_{\zeta',-\bq}}-\ev*{a^\dagger_{\zeta,\bq}}\ev*{a^\dagger_{\zeta',-\bq}}$ is the correlated part of the two-photon expectation value. To project this equation of motion onto the 1s-exciton subspace, we multiply by $\phi_{\bk-\beta\bq}$, sum over $\bk$, and express the two-pair expectation value in terms of $\tilde{\mathcal{B}}^{\zeta\zeta'}_{\bp,\pm}$ to obtain
\begin{align}
  -i\hbar\partial_t\mathcal{C}^{\zeta\zeta'}_\bq &=
  (E^{\rm p}_\bq + E^{\rm x}_{\bq})\mathcal{C}^{\zeta\zeta'}_\bq + \Omega_{-\bq}^*\mathcal{D}^{\zeta\zeta'}_{\bq}
  - \delta_{\zeta,\zeta'}\tilde{\Omega}'_\bq \ev*{P^\dagger_{\zeta,0}}\ev*{P^\dagger_{\zeta,0}}
  + \tilde{A}_{\bq',\bq}^\pm \tilde{\mathcal{B}}^{\zeta\zeta'}_{-\bq',\pm},
\end{align}
where
$
\tilde{\Omega}'_{\bq} =
A_\bq\sum_{\bk}
\phi_{\bk-\beta\bq}
\phi^{*}_{\bk}
\phi^{*}_{\bk-\bq}
$
and
$
\tilde{A}_{\bq',\bq}^\pm
= \tilde{\Omega}_{\bq}\delta_{\bq\bq'}
  \mp A_\bq
  \sum_{\bk}\phi_{\bk + \alpha\bq}
  \phi_{\bk+\bq-\beta\bq'}^*
  \phi_{\bk-\alpha\bq'}^*
$.

The equation of motion for the correlated two-photon amplitude $\mathcal{D}^{\zeta\zeta'}_\bq$ is derived in a similar manner and takes the form
\begin{align}
-i\hbar\partial_t \mathcal{D}^{\zeta\zeta'}_\bq &=
2\tilde{E}^{\rm p}_\bq
\mathcal{D}^{\zeta\zeta'}_\bq
+ \Omega_\bq \mathcal{C}^{\zeta'\zeta}_{-\bq}
+ \Omega_{-\bq}\mathcal{C}^{\zeta\zeta'}_\bq.
\end{align}

For the biexcitonic correlations, we start out with the equation of motion of the correlated two-pair amplitude, which takes the form
\begin{align}
\label{eq:two-pair-correlation-eom}
  \begin{split}
  -i\hbar\partial_t \ev*{c^\dagger_{\zeta_1,\bk_1+\bq}v_{\zeta_1',\bk_1}c^\dagger_{\zeta_2,\bk_2-\bq}v_{\zeta_2',\bk_2}}^{\rm c} & =
  \delta_{\zeta_1,\zeta_1'} A^*_{\bq}
  \ev*{a_{\zeta_1,\bq}^\dagger c^\dagger_{\zeta_2,\bk_2-\bq} v_{\zeta_2',\bk_2}}^{\rm c}
  + \delta_{\zeta_2,\zeta_2'} A^*_{-\bq}
  \ev*{a_{\zeta_2,-\bq}^\dagger c_{\zeta_1,\bk_1+\bq}^\dagger v_{\zeta_1',\bk_1}}^{\rm c}
  \\
  &-\delta_{\zeta_1',\zeta_2} A^*_{\bk_2-\bk_1-\bq}
  \ev*{a_{\zeta_2,\bk_2-\bk_1-\bq}^\dagger c^\dagger_{\zeta_1,\bk_1+\bq} v_{\zeta_2',\bk_2}}^{\rm c}
  - \delta_{\zeta_1,\zeta_2'} A^*_{\bk_1+\bq-\bk_2}
  \ev*{a_{\zeta_1,\bk_1+\bq-\bk_2}^\dagger c_{\zeta_2,\bk_2-\bq}^\dagger v_{\zeta_1',\bk_1}}^{\rm c}
  \\
  &+V_\bq
    \qty[\ev*{c_{\zeta_1,\bk_1}^\dagger v_{\zeta_1',\bk_1}}
    -\ev*{c_{\zeta_1,\bk_1+\bq}^\dagger v_{\zeta_1',\bk_1+\bq}}]
    \qty[\ev*{c_{\zeta_2,\bk_2}^\dagger v_{\zeta_2',\bk_2}}
   -\ev*{c_{\zeta_2,\bk_2-\bq}^\dagger v_{\zeta_2',\bk_2-\bq}}]
  \\
  &-V_{\bk_2-\bk_1-\bq}
  \qty[\ev*{c_{\zeta_2,\bk_1}^\dagger v_{\zeta_1',\bk_1}}
  -\ev*{c_{\zeta_2,\bk_2-\bq}^\dagger v_{\zeta_1',\bk_2-\bq}}]
  \qty[\ev*{c_{\zeta_1,\bk_2}^\dagger v_{\zeta_2',\bk_2}} -
    \ev*{c_{\zeta_1,\bk_1+\bq}^\dagger v_{\zeta_2',\bk_1+\bq}}]
  \\
  &+ (E^{\rm c}_{\bk_1+\bq} + E^{\rm c}_{\bk_2-\bq}
  - E^{\rm v}_{\bk_1} - E^{\rm v}_{\bk_2})
  \ev*{c^\dagger_{\zeta_1,\bk_1+\bq}v_{\zeta_1',\bk_1}c^\dagger_{\zeta_2,\bk_2-\bq}v_{\zeta_2',\bk_2}}^{\rm c}
  \\
  &+ \sum_\bp V_\bp\Big[
    \ev*{c^\dagger_{\zeta_1,\bk_1+\bq+\bp} v_{\zeta_1',\bk_1}
         c_{\zeta_2,\bk_2-\bq-\bp}^\dagger v_{\zeta_2',\bk_2}}^{\rm c}
  + \ev*{c^\dagger_{\zeta_1,\bk_1+\bq} v_{\zeta_1',\bk_1-\bp}
         c_{\zeta_2,\bk_2-\bq}^\dagger v_{\zeta_2',\bk_2+\bp}}^{\rm c}
  \\
  &\hspace{1.5cm}
  - \ev*{c^\dagger_{\zeta_1\bk_1+\bq+\bp}v_{\zeta_1'\bk_1}
         c_{\zeta_2\bk_2-\bq}^\dagger v_{\zeta_2'\bk_2+\bp}}^{\rm c}
  - \ev*{c^\dagger_{\zeta_1\bk_1+\bq} v_{\zeta_1'\bk_1-\bp}
         c_{\zeta_2\bk_2-\bq-\bp}^\dagger v_{\zeta_2'\bk_2}}^{\rm c}
  \\
  &\hspace{1.5cm}
  - \ev*{c_{\zeta_1\bk_1+\bq+\bp}^\dagger v_{\zeta_1'\bk_1+\bp}
         c_{\zeta_2\bk_2-\bq}^\dagger v_{\zeta_2',\bk_2}}^{\rm c}
  - \ev*{c_{\zeta_1\bk_1+\bq}^\dagger v_{\zeta_1'\bk_1}
         c_{\zeta_2\bk_2-\bq+\bp}^\dagger v_{\zeta_2',\bk_2+\bp}}^{\rm c}\Big].
  \\
  \end{split}
  \end{align}
To obtain the equation of motion for $\tilde{\mathcal{B}}^{\zeta\zeta'}_{\bq,\pm}$ from this, we use the relation from Eq.~\eqref{eq:B-tilde-expansion} as
\begin{align}
\label{eq:sing-trip-corr}
  \begin{split}
\frac{1}{2}&\qty(
\ev*{c_{\zeta_1,\bk_1+\bq}^\dagger v_{\zeta_1,\bk_1}
     c_{\zeta_2,\bk_2-\bq}^\dagger v_{\zeta_2,\bk_2}}^{\rm c}
\pm
\ev*{c_{\zeta_2,\bk_1+\bq}^\dagger v_{\zeta_1,\bk_1}
     c_{\zeta_1,\bk_2-\bq}^\dagger v_{\zeta_2,\bk_2}}^{\rm c})
\\ &=
\sum\Big\{
\phi_{\bk_1 + \beta\bq}^*
\phi_{\bk_2 - \beta\bq}^*
\tilde{\mathcal{B}}_{\bq,\pm}^{\zeta_1\zeta_2}
\mp
\phi_{\alpha\bk_1 + \beta(\bk_2-\bq)}^*
\phi_{\beta(\bk_1+\bq) + \alpha\bk_2}^*
\tilde{\mathcal{B}}_{\bk_2-\bk_1-\bq,\pm}^{\zeta_1\zeta_2}.
\Big\}
\end{split}
\end{align}
Inserting Eq.~\eqref{eq:two-pair-correlation-eom} into Eq.~\eqref{eq:sing-trip-corr} and subsequently multiplying by $\phi_{\bk_1+\beta\bq}\phi_{\bk_2-\beta\bq}$ and summing over $\bk_1,\bk_2$, we find
\begin{align}
\label{eq:two-pair-correlation-eom-2}
\begin{split}
-i\hbar
\sum_{\bq'}
\mathcal{S}_{\bq,\bq'}^\pm
\partial_t \tilde{\mathcal{B}}_{\bq'}^{\zeta_1\zeta_2\pm}
&=
\sum_{\bq'}H_{\bq,\bq'}^{\pm}
B_{\bq',\pm}^{\zeta_1\zeta_2}
+
\frac{1}{2}(1\pm\delta_{\zeta_1\zeta_2})
\sum_{\bq'}\qty[
\tilde{A}_{-\bq,-\bq'}^{\pm*}
\mathcal{C}^{\zeta_2\zeta_1}_{-\bq'}
+
\tilde{A}_{\bq,\bq'}^{\pm*}
\mathcal{C}^{\zeta_1\zeta_2}_{\bq'}
]
+\frac{1}{2}(1\pm\delta_{\zeta_1\zeta_2})\tilde{W}_{\bq,0}^{\pm}
\ev*{P_{\zeta_1,0}^\dagger}
\ev*{P_{\zeta_2,0}^\dagger},
\end{split}
\end{align}
where
$
\mathcal{S}_{\bq,\bq'}^\pm  =
  \delta_{\bq\bq'}
  \mp
  \sum_{\bk}
  \phi_{\bk-\alpha\bq}
  \phi_{\bk+\bq'-\beta\bq}
  \phi_{\bk-\bq+\beta\bq'}^*
  \phi_{\bk+\alpha\bq'}^*
  $
is an exciton wavefunction overlap matrix and $H_{\bq,\bq'}^\pm$ is the homogeneous part of the equation of motion, given by
\begin{align}
\label{eq:bix-hamiltonian}
  H^\pm_{\bq,\bq'} = \mathcal{S}_{\bq,\bq'}^\pm\qty(2E^{\rm x}_0 + \frac{\hbar^2 q^{\prime 2}}{M}) + \tilde{W}^\pm_{\bq,\bq'},
\end{align}
with $M=m_{\rm e}+m_{\rm h}$ the total exciton mass and $\tilde{W}^{\pm}_{\bq,\bq'}$ an exciton-exciton Coulomb scattering matrix,
\begin{align}
  \begin{split}
    \tilde{W}_{\bq,\bq'}^\pm
    &=\sum_{\bk_1\bk_2} V_{\bq'-\bq}
    \phi_{\bk_1}
    \phi_{\bk_2}
    \Big[
    \phi_{\bk_1-\beta(\bq-\bq')}^*
    -\phi_{\bk_1+\alpha(\bq-\bq')}^*
    \Big]
    \Big[
    \phi_{\bk_2+\beta(\bq-\bq')}^*
    -\phi_{\bk_2-\alpha(\bq-\bq')}^*
    \Big]
    \\
    &\pm
    \sum_{\bk_1\bk_2}V_{\bk_1-\bk_2+(\alpha-\beta)\bq+\bq'}
    \phi_{\bk_1}\phi_{\bk_2}
    \Big[
    \phi_{\bk_1-\beta(\bq-\bq')}^*
    -\phi_{\bk_2-\alpha\bq-\alpha\bq'}^*
    \Big]
    \Big[
    \phi_{\bk_1+\alpha\bq+\alpha\bq'}^*
    -\phi_{\bk_2+\beta(\bq-\bq')}^*
    \Big].
  \end{split}
\end{align}
Notice that the single-momentum Coulomb matrix $\tilde{W}^\pm_{\bq}$ in Eqs.~\eqref{eq:exciton-eom} and~\eqref{eq:exciton-eom-coefficients} is simply shorthand for $\tilde{W}^\pm_{\bq}=\tilde{W}^{\pm}_{\bq,0}$.

Since the equation of motion Eq.~\eqref{eq:two-pair-correlation-eom-2} is not momentum-diagonal, i.e. it couples $\tilde{\mathcal{B}}^{\zeta\zeta'}_{\bq,\pm}$ with $\tilde{\mathcal{B}}^{\zeta\zeta'}_{\bq',\pm}$, it is advantageous to transform to a diagonalised basis, $\mathcal{B}^{\zeta\zeta'}_{\mu,\pm}$ via the biexcitonic wavefunctions $\Phi^{\pm}_{\mu,\bq}$ as $\tilde{\mathcal{B}}^{\zeta\zeta'}_{\bq,\pm}=\sum_\mu \Phi^{\pm}_{\mu,\bq} \mathcal{B}^{\zeta\zeta'}_{\mu,\pm}$. The biexcitonic wavefunction $\Phi^{\pm}_{\mu,\bq}$ is then defined as the solution to the eigenvalue equation
\begin{align}
\label{eq:bix-eigenvalue}
  \sum_{\bq'\bq''}(\mathcal{S}^\pm)^{-1}_{\bq,\bq'} H_{\bq',\bq''}^\pm \Phi^\pm_{\mu,\bq''} = E^{\rm xx}_{\mu,\pm}\Phi^\pm_{\mu,\bq},
\end{align}
where $\mu$ is an index that labels the biexcitonic eigenstates. Since Eq.~\eqref{eq:bix-eigenvalue} is non-Hermitian, a dual set of wavefunctions $\widebar{\Phi}^{\pm}_{\mu,\bq}$ must be explicitly defined from the orthogonality relation $\sum_{\bq}\widebar{\Phi}^{\pm}_{\mu,\bq}\Phi^{\pm}_{\mu',\bq} = \delta_{\mu,\mu'}$. If we had defined the biexcitonic expansion slightly different as $\sum_{\bq'} (\mathcal{S}^{\pm})^{-1/2}_{\bq,\bq'} \tilde{\mathcal{B}}^{\zeta\zeta'}_{\bq',\pm}=\sum_\mu \Phi^{\pm}_{\mu,\bq'}  \mathcal{B}^{\zeta\zeta'}_{\mu,\pm}$, we would have ended up with a Hermitian eigenvalue equation~\cite{schafer2013semiconductorSI}. However, here we retain the non-Hermitian property for computational simplicity. The time evolution of the other variables is necessarily unaffected by this choice.

Writing the biexcitonic amplitudes in terms of the diagonalised basis, the equations of motion reduce to
\begin{align}
\label{eq:full-eom}
\begin{split}
  -i\hbar\partial_t \ev*{a_{\zeta,0}^\dagger} &= E^{\rm p}_0\ev*{a_{\zeta,0}^\dagger} + \Omega_0 \ev*{P^\dagger_{\zeta,0}}
  +i\hbar\sqrt{2\gamma^{\rm p}}\!\ev*{a^{\rm in\dagger}_{\zeta}}
  \\
  -i\hbar\partial_t\ev*{P_{\zeta,0}^\dagger} &= E^{\rm x}_0\ev*{P^{\dagger}_{\zeta,0}} + \Omega_0\ev*{a_{\zeta,0}^\dagger}
  -
  \sum_\bq \tilde{\Omega}_\bq\qty(\mathcal{C}^{\zeta\zeta'}_\bq + \delta_{\bq,0}\ev*{a_{\zeta,0}^\dagger}\ev*{P_{\zeta,0}^\dagger})\ev*{P_{\zeta,0}}
  + W^0 \abs*{\ev*{P_{\zeta,0}^\dagger}}^2\ev*{P_{\zeta,0}^\dagger}
  +\sum_{\mu\zeta'\pm} W_{\mu}^{\pm}
  \mathcal{B}_{\mu,\pm}^{\zeta\zeta'} \ev*{P_{\zeta',0}}.
\\
-i\hbar\partial_t \mathcal{B}_{\mu,\pm}^{\zeta\zeta'} &= E^{\rm xx}_{\mu,\pm}\mathcal{B}_{\mu,\pm}^{\zeta\zeta'}
+
\frac{1}{2}(1\pm\delta_{\zeta\zeta'})
\{\widebar{W}^\pm_{\mu}
\ev*{P^\dagger_{\zeta,0}}\ev*{P^\dagger_{\zeta',0}}
+
\sum_{\bq}
[
\widebar{\Omega}_{\mu,-\bq}^\pm
\mathcal{C}^{\zeta'\zeta}_{-\bq}
+
\widebar{\Omega}_{\mu,\bq}^\pm
\mathcal{C}^{\zeta\zeta'}_\bq
]
\}
\\
-i\hbar\partial_t \mathcal{C}^{\zeta\zeta'}_\bq &=
(E^{\rm p}_{\bq}+E^{\rm x}_{\bq})
 \mathcal{C}^{\zeta\zeta'}_\bq
 + \Omega_{\bq} \mathcal{D}^{\zeta\zeta'}_{\bq}
 - \frac{1}{2}\delta_{\zeta\zeta'}\tilde{\Omega}_\bq \ev*{P^\dagger_{\zeta,0}}^2
 + \sum_{\mu\pm} \Omega_{\mu,\bq}^{\pm}\mathcal{B}_{\mu,\pm}^{\zeta\zeta'}
\\
-i\hbar\partial_t \mathcal{D}^{\zeta\zeta'}_\bq &=
2E^{\rm p}_\bq
\mathcal{D}^{\zeta\zeta'}_\bq
+ \Omega_\bq \mathcal{C}^{\zeta'\zeta}_{-\bq}
+ \Omega_{-\bq}\mathcal{C}^{\zeta\zeta'}_\bq
.
\end{split}
\end{align}
with the biexcitonic coefficients
\begin{align}
  \begin{split}
    W^\pm_\mu &= \sum_\bq \Phi^\pm_{\mu,\bq} \tilde{W}^{\pm*}_{\bq,0},
    \hspace{0.5cm}
    \widebar{W}^\pm_\mu = \sum_{\bq\bq'} \widebar{\Phi}^\pm_{\mu,\bq}
    (\mathcal{S}^\pm)^{-1}_{\bq,\bq'}\tilde{W}^\pm_{\bq',0}
    \\
    \Omega^\pm_{\mu,\bq} &= \sum_{\bq'} \Phi^\pm_{\mu,-\bq'}\tilde{A}^\pm_{\bq',\bq},
    \hspace{0.5cm}
    \widebar{\Omega}^\pm_{\mu,\bq} = \sum_{\bq'\bq''}\widebar{\Phi}^\pm_{\mu,\bq'}
    (\mathcal{S}^{\pm})^{-1}_{\bq',\bq''} \tilde{A}^{\pm*}_{\bq'',\bq}.
  \end{split}
\end{align}
Since the 1s-exciton wavefunction and the electron-photon coupling strength $A_\bq$ can be taken real without loss of generality, we take all of matrix elements in the equation of motion to be real. With this, we also have $\tilde{\Omega}'_\bq = \frac{1}{2}\tilde{\Omega}_\bq$.

We take the input field to be monochromatic, such that $\ev*{a^{\rm in\dagger}_{\zeta}} = \ev*{\hat{a}^{\rm in\dagger}_{\zeta}}e^{i\omega_{\rm d}t}$, where $\ev*{\hat{a}^{\rm in\dagger}_{\zeta}}$ is constant and $\omega_{\rm d}$ is the driving frequency. We then transform to a rotating reference frame with respect to $\omega_{\rm d}$ by introducing the slowly-varying dynamical variables
$
\ev*{\hat{a}^\dagger_{\zeta,0}} = \ev*{a^\dagger_{\zeta,0}}e^{-i\omega_{\rm d}t},\;
\ev*{\hat{P}^\dagger_{\zeta,0}} = \ev*{P^\dagger_{\zeta,0}}e^{-i\omega_{\rm d}t},\;
\hat{\mathcal{B}}^{\zeta\zeta'}_{\mu,\pm} = \mathcal{B}^{\zeta\zeta'}_{\mu,\pm} e^{-2i\omega_{\rm d}t}, \;
\hat{\mathcal{C}}^{\zeta\zeta'}_\bq = \mathcal{C}^{\zeta\zeta'}_\bq e^{-2i\omega_{\rm d}t},\;
\hat{\mathcal{D}}^{\zeta\zeta'}_\bq = \mathcal{D}^{\zeta\zeta'}_\bq e^{-2i\omega_{\rm d}t}
$.
The equation of motion for these are identical to Eq.~\eqref{eq:full-eom} with the substitutions
$
\ev*{a^{\rm in\dagger}}\rightarrow\ev*{\hat{a}^{\rm in\dagger}}, \;
E^{\rm p}_\bq\rightarrow E^{\rm p}_{\bq}-\hbar\omega_{\rm d},\;
E^{\rm x}_\bq\rightarrow E^{\rm x}_{\bq}-\hbar\omega_{\rm d}, \;
E^{\rm xx}_{\mu,\pm}\rightarrow E^{\rm xx}_{\mu,\pm}-2\hbar\omega_{\rm d}
$. Thus, the carets can simply be dropped. By introducing the phonon-induced broadening of the exciton and biexciton energies $\tilde{E}^{\rm x}_\bq,\: \tilde{E}^{\rm xx}_{\mu,\pm}$ and the broadening of the photon energy $\tilde{E}^{\rm p}_{\bq}$ as described in the main text, Eq.~\eqref{eq:full-eom} becomes Eq.~(1) of the main text.

\section{Heisenberg-Langevin equations for fluctuation operators}
\label{sec:heisenberg-langevin}
The equations of motion for the fluctuation operators are derived in a similar manner to the expectation values. Due to the fluctuation-dissipation theorem, the broadening of the energy levels $\gamma^{\rm x}$ and $\gamma^{\rm p}$ must be accompanied by Langevin noise terms in the equations of motion for the operators, such that the commutation relations are preserved~\cite{lax1966quantumSI}. We implement these noise sources at the level of the photon and electron-hole pair operators. For the photon fluctuations $\delta a_{\zeta,0}^\dagger = a^\dagger_{\zeta,0}-\ev*{a^\dagger_{\zeta,0}}$, we have (in the rotating frame)
\begin{align}
\label{eq:photon-fluctuation-eom}
  -i\hbar\partial_t\delta a^\dagger_{\zeta,0} =
  (\tilde{E}^{\rm p}_0-\hbar\omega_{\rm d}) \delta a^\dagger_{\zeta,0}
  + \Omega_0\delta P^\dagger_{\zeta,0} + i\hbar\sqrt{2\gamma^{\rm p}}\delta a^{\rm in\dagger}_{\zeta},
\end{align}
where $\delta P^\dagger_{\zeta,0} =  P^\dagger_{\zeta,0} - \ev*{P^\dagger_{\zeta,0}}$ and the input fluctuation field $\delta a^{\rm in\dagger}_{\zeta}$ has the properties~\cite{lax1966quantumSI}
\begin{align}
  \ev*{\delta a^{\rm in}_{\zeta}} = \ev*{\delta a^{\rm in\dagger}_{\zeta}(t)\delta a^{\rm in}_{\zeta'}(t')} = \ev*{\delta a^{\rm in}_{\zeta}(t)\delta a^{\rm in}_{\zeta'}(t')} = 0,
  \;\;\;\;
  \ev*{\delta a^{\rm in}_{\zeta}(t)\delta a^{\rm in\dagger}_{\zeta'}(t')} = \delta_{\zeta,\zeta'}\delta(t-t').
\end{align}
We note that the photon Langevin noise term can be derived explicitly from the microscopic interactions between the cavity and the electromagnetic environment~\cite{steck2007quantumSI,gardiner2004quantumSI} or within a quasi-normal mode expansion of the electric field operator~\cite{franke2019quantizationSI}.

For the exciton fluctuations, we first derive the electron-hole pair fluctuations $\delta(c^\dagger_{\zeta,\bk} v_{\zeta,\bk}) = c^\dagger_{\zeta,\bk} v_{\zeta,\bk} - \ev*{c^\dagger_{\zeta,\bk} v_{\zeta,\bk}}$ (in the rotating frame),
\begin{align}
\label{eq:pair-fluctuation-eom}
  \begin{split}
    -i\hbar\partial_t \delta(c^\dagger_{\zeta,\bk} v_{\zeta,\bk})
    &= i\hbar\sqrt{2\gamma^{\rm x}} F_{\bk,\bk}^{\dagger\zeta\zeta} + A^*_0 \delta a^\dagger_{\zeta,0}
  \\
  &- \sum_{\zeta_1\bk_1\bq}
  \Big\{ A^*_{\bq} [
    a_{\zeta,\bq}^\dagger c^\dagger_{\zeta,\bk} v_{\zeta_1,\bk_1}
    v^\dagger_{\zeta_1,\bk_1} c_{\zeta,\bk+\bq}
  - \ev*{a_{\zeta,\bq}^\dagger c^\dagger_{\zeta\bk} v_{\zeta_1\bk_1}
    v^\dagger_{\zeta_1\bk_1} c_{\zeta\bk+\bq}}]
  \\ &\hspace{2cm} +
   A^*_{-\bq}[
   a_{\zeta,-\bq}^\dagger c^\dagger_{\zeta_1,\bk_1} v_{\zeta,\bk}
   v^\dagger_{\zeta,\bk+\bq}c_{\zeta_1,\bk_1}
   -\ev*{a_{\zeta,-\bq}^\dagger c^\dagger_{\zeta_1,\bk_1} v_{\zeta,\bk}
    v^\dagger_{\zeta,\bk+\bq}c_{\zeta_1,\bk_1}}]
  \Big\}
  \\ &+
  (E^{\rm c}_\bk-E^{\rm v}_\bk -\hbar\omega_{\rm d} + i\gamma^{\rm x})
  \delta(c^\dagger_{\zeta,\bk}v_{\zeta,\bk})
  - \sum_\bq V_\bq \delta(c_{\zeta,\bk+\bq}^\dagger v_{\zeta,\bk+\bq})
  \\
  &+ \sum_{\zeta_1\bk_1}\sum_{\zeta_2\bk_2}\sum_\bq V_\bq\Big\{\big[
    c^\dagger_{\zeta,\bk+\bq}v_{\zeta,\bk}c_{\zeta_1,\bk_1-\bq}^\dagger v_{\zeta_2,\bk_2}
+   c^\dagger_{\zeta,\bk}v_{\zeta,\bk+\bq}c_{\zeta_1,\bk_1}^\dagger v_{\zeta_2,\bk_2-\bq}
\big]
\big[
v_{\zeta_2,\bk_2}^\dagger c_{\zeta_1,\bk_1}
- v_{\zeta_2,\bk_2-\bq}^\dagger c_{\zeta_1,\bk_1-\bq}
\big]
\\
&\hspace{2.5cm}
-\ev{
\big[
  c^\dagger_{\zeta,\bk+\bq}v_{\zeta,\bk}c_{\zeta_1,\bk_1-\bq}^\dagger v_{\zeta_2,\bk_2}
+   c^\dagger_{\zeta,\bk}v_{\zeta,\bk+\bq}c_{\zeta_1,\bk_1}^\dagger v_{\zeta_2,\bk_2-\bq}
\big]
\big[
v_{\zeta_2,\bk_2}^\dagger c_{\zeta_1,\bk_1}
- v_{\zeta_2,\bk_2-\bq}^\dagger c_{\zeta_1,\bk_1-\bq}
\big]
}
\Big\},
  \end{split}
\end{align}
where $F^{\zeta\zeta'}_{\bk,\bk'}$ is the Langevin noise operator for the electron-hole pair with the properties~\cite{lax1966quantumSI} $\ev*{F^{\zeta_1\zeta_2}_{\bk_1,\bk_2}}=\ev*{F^{\zeta_1\zeta_2}_{\bk_1,\bk_2}(t)F^{\zeta_1'\zeta_2'}_{\bk_1',\bk_2'}(t')}=0,\; \ev*{F^{\zeta_1\zeta_2}_{\bk_1,\bk_2}(t)F^{\dagger\zeta_1'\zeta_2'}_{\bk_1',\bk_2'}(t')} = \delta(t-t')\delta_{\bk_1,\bk_1}\delta_{\bk_2,\bk_2'}\delta_{\zeta_1,\zeta_1'}\delta_{\zeta_2,\zeta_2'}$
. Whereas the photon Langevin noise source can be microscopically derived, the corresponding noise term for electrons and holes is introduced phenomenologically to counterbalance the dephasing by following the general procedure in Ref.~\cite{lax1966quantumSI} that ensures conservation of the commutation relation and respects the fluctuation-dissipation theorem.

 Using the factorization rules for the expectation values as described below Eq.~\eqref{eq:unit-operator-expansion}, we can decompose the five-operator fluctuations in terms of two- and three-particle fluctuations as
\begin{align}
\label{eq:acv-fluctuation}
\begin{split}
a_{\zeta,\bq}^\dagger c^\dagger_{\zeta,\bk} v_{\zeta_1,\bk_1}
v^\dagger_{\zeta_1,\bk_1} c_{\zeta,\bk+\bq}
- \ev*{a_{\zeta,\bq}^\dagger c^\dagger_{\zeta\bk} v_{\zeta_1\bk_1}
v^\dagger_{\zeta_1\bk_1} c_{\zeta\bk+\bq}}
&= \delta(a_{\zeta,\bq}^\dagger c^\dagger_{\zeta,\bk} v_{\zeta_1,\bk_1})
\ev*{v^\dagger_{\zeta_1,\bk_1} c_{\zeta,\bk+\bq}}
\\ &+ \qty[\delta(a_{\zeta,\bq}^\dagger c^\dagger_{\zeta,\bk} v_{\zeta_1,\bk_1})
+ \ev*{a_{\zeta,\bq}^\dagger c^\dagger_{\zeta,\bk} v_{\zeta_1,\bk_1}}]
\delta(v^\dagger_{\zeta_1,\bk_1} c_{\zeta,\bk+\bq}),
\end{split}
\end{align}
where $\delta(a_{\zeta,\bq}^\dagger c^\dagger_{\zeta,\bk} v_{\zeta_1,\bk_1}) :=
a_{\zeta,\bq}^\dagger c^\dagger_{\zeta,\bk} v_{\zeta_1,\bk_1} - \ev*{a_{\zeta,\bq}^\dagger c^\dagger_{\zeta,\bk} v_{\zeta_1,\bk_1}}$ is a three-particle electron-hole-photon fluctuation operator.
Similarly, the six-operator fluctuations can be decomposed in terms of two- and four-particle fluctuations as
\begin{align}
\label{eq:cvcv-fluctuation}
\begin{split}
c^\dagger_{\zeta,\bk+\bq}v_{\zeta,\bk}
c^\dagger_{\zeta_1,\bk_1-\bq} v_{\zeta_2,\bk_2}
v_{\zeta_2,\bk_2}^\dagger c_{\zeta_1,\bk_1}
-
\ev*{c^\dagger_{\zeta,\bk+\bq}v_{\zeta,\bk}
c^\dagger_{\zeta_1,\bk_1-\bq} v_{\zeta_2,\bk_2}
v_{\zeta_2,\bk_2}^\dagger c_{\zeta_1,\bk_1}}
&=
\delta(c^\dagger_{\zeta,\bk+\bq}v_{\zeta,\bk}
c^\dagger_{\zeta_1,\bk_1-\bq} v_{\zeta_2,\bk_2})
\ev*{v_{\zeta_2,\bk_2}^\dagger c_{\zeta_1,\bk_1}}
\\ &\hspace{-4cm}+
\qty[\delta(c^\dagger_{\zeta,\bk+\bq}v_{\zeta,\bk}
c^\dagger_{\zeta_1,\bk_1-\bq} v_{\zeta_2,\bk_2}) +
\ev*{c^\dagger_{\zeta,\bk+\bq}v_{\zeta,\bk}
c^\dagger_{\zeta_1,\bk_1-\bq} v_{\zeta_2,\bk_2}}]
\delta(v_{\zeta_2,\bk_2}^\dagger c_{\zeta_1,\bk_1}),
\end{split}
\end{align}
where $\delta(c^\dagger_{\zeta,\bk+\bq}v_{\zeta,\bk}
c^\dagger_{\zeta_1,\bk_1-\bq} v_{\zeta_2,\bk_2}) := c^\dagger_{\zeta,\bk+\bq}v_{\zeta,\bk}
c^\dagger_{\zeta_1,\bk_1-\bq} v_{\zeta_2,\bk_2} - \ev*{c^\dagger_{\zeta,\bk+\bq}v_{\zeta,\bk}
c^\dagger_{\zeta_1,\bk_1-\bq} v_{\zeta_2,\bk_2}}$ is a two-electron-hole-pair fluctuation operator. The decomposition in Eqs.~\eqref{eq:acv-fluctuation} and~\eqref{eq:cvcv-fluctuation} follows directly from DCT factorization rules of the six- and five-operator expectation values as described below Eq.~\eqref{eq:unit-operator-expansion}.

Assuming that the fluctuations are small, we perform a linearization of the equations of motion of the fluctuations with respect to the one-, two-, three- and four-particle fluctuation operators as introduced. This means that the terms in Eqs.~\eqref{eq:acv-fluctuation} and \eqref{eq:cvcv-fluctuation} involving products of fluctuation operators are discarded.

The electron-hole-photon fluctuation operator are projected onto the 1s-exciton subspace in order to define the exciton-photon fluctuation operator $\delta\mathcal{C}^{\zeta\zeta'}_{\bq}$ as
\begin{align}
  \delta\mathcal{C}^{\zeta\zeta'}_{\bq} = \sum_\bk \phi_{\bk}
  \delta(a^\dagger_{\zeta,\bq} c^\dagger_{\zeta',\bk-\alpha\bq}v_{\zeta',\bk+\beta\bq}).
\end{align}
Similarly, we can define the biexcitonic fluctuation operator $\delta\tilde{\mathcal{B}}^{\zeta\zeta'}_{\bq,\pm}$ through the relation
\begin{align}
  \begin{split}
  \delta(c^\dagger_{\zeta_1,\bk_1+\bq}v_{\zeta_1,\bk_1}c_{\zeta_2\bk_2}^\dagger v_{\zeta_2\bk_2+\bq}) &= \sum_\pm\qty[
  \phi^*_{\bk_1+\beta\bq}
  \phi^*_{\bk_2+\alpha\bq}
  \delta \tilde{\mathcal{B}}_{\bq,\pm}^{\zeta_1\zeta_2}
   \mp  \phi^*_{\beta\bk_2+\alpha\bk_1}
  \phi^*_{\beta(\bk_1+\bq)+\alpha(\bk_2+\bq)}
  \delta \tilde{\mathcal{B}}_{\bk_2-\bk_1,\pm}^{\zeta_1\zeta_2}.
  ]
  \end{split}
\end{align}
The biexcitonic fluctuation operators $\delta\tilde{\mathcal{B}}_{\bq,\pm}^{\zeta\zeta'}$ are expanded on the biexcitonic wavefunctions as $\delta\tilde{\mathcal{B}}^{\zeta\zeta'}_{\bq,\pm}=\sum_\mu\Phi^{\pm}_{\mu,\bq}\delta \mathcal{B}^{\zeta\zeta'}_{\mu,\pm}$.

In addition to $\delta\mathcal{B}$ and $\delta\mathcal{C}$, we also define the two-photon fluctuation operator $\delta\mathcal{D}^{\zeta\zeta'}_\bq = \delta(a^\dagger_{\zeta,\bq}a^\dagger_{\zeta',-\bq}) := a^\dagger_{\zeta,\bq}a^\dagger_{\zeta',-\bq} - \ev*{a^\dagger_{\zeta,\bq}a^\dagger_{\zeta',-\bq}}$.

We now project Eq.~\eqref{eq:pair-fluctuation-eom} onto the 1s-exciton subspace by multiplying by $\phi_{\bk}$ and summing over $\bk$, and impose the fluctuation linearization and expressing the electron-hole-photon and two-pair fluctuations in terms of $\delta\mathcal{C}^{\zeta\zeta'}_{\bq}$ and $\delta\mathcal{\mathcal{B}}^{\zeta\zeta'}_{\mu,\pm}$, leading to
\begin{align}
\label{eq:exciton-fluctuation-eom}
  \begin{split}
    -i\hbar\partial_t \delta P_{\zeta,0}^\dagger
    &= (\tilde{E}_0^{\rm x} - \hbar\omega_{\rm d}) \delta P_{\zeta,0}^\dagger
    + \Omega_0\delta a_{\zeta,0}^\dagger
    -\sum_\bq \tilde{\Omega}_\bq
    \qty[
    \delta\mathcal{C}^{\zeta\zeta}_\bq
    \ev*{P_{\zeta, 0}}
    +
    \qty(\delta_{\bq,0}\ev*{a_{\zeta,0}^\dagger}\ev*{P_{\zeta,0}^\dagger}
    + \mathcal{C}^{\zeta\zeta}_\bq)
    \delta P_{\zeta, 0}]
    \\
    &+W^0
    \ev*{P^\dagger_{\zeta,0}}^2
    \delta P_{\zeta,0}
    +
    \sum_{\zeta'\mu\pm}
    W^\pm_\mu
    \qty[\delta\mathcal{B}_{\mu,\pm}^{\zeta\zeta'}\ev*{P_{\zeta',0}}
    +\mathcal{B}_{\mu,\pm}^{\zeta\zeta'} \delta P_{\zeta',0}]
    + i\hbar\sqrt{2\gamma^{\rm x}} \delta P_{\zeta,0}^{\rm in\dagger},
  \end{split}
\end{align}
where $\delta P^{\rm in\dagger}_{\zeta,\bq}=\sum_{\bk}\phi_{\bk}F^{\dagger\zeta\zeta}_{\bk+\alpha\bq,\bk-\beta\bq}$ is the exciton Langevin noise operator with properties $\ev*{\delta P^{\rm in}_{\zeta,\bq}(t)}=\ev*{\delta P^{\rm in\dagger}_{\zeta,\bq}(t)\delta P^{\rm in}_{\zeta',\bq'}(t')}=\ev*{\delta P^{\rm in}_{\zeta,\bq}(t)\delta P^{\rm in}_{\zeta',\bq'}(t')}=0,\;\;\ev*{\delta P^{\rm in}_{\zeta,\bq}(t)\delta P^{\rm in\dagger}_{\zeta',\bq'}(t')}=\delta(t-t')\delta_{\zeta,\zeta'}\delta_{\bq,\bq'}$.

The derivation of the linearised Heisenberg-Langevin equations for the multiparticle fluctuations $\delta\mathcal{B},\;\delta\mathcal{C}$ and $\delta\mathcal{D}$ is analogous to the derivation presented in Sec.~\ref{sec:dct-eom} with the result
\begin{align}
\label{eq:multiparticle-fluctuation-eom}
\begin{split}
-i\hbar\partial_t \delta\mathcal{B}_{\mu,\pm}^{\zeta\zeta'} &= (\tilde{E}^{\rm xx}_{\mu,\pm}-2\hbar\omega_{\rm d})\delta\mathcal{B}_{\mu,\pm}^{\zeta\zeta'}
+
\frac{1}{2}(1\pm\delta_{\zeta\zeta'})
\qty[
\widebar{\Omega}_{\mu,-\bq}^\pm
\delta\mathcal{C}^{\zeta'\zeta}_{-\bq}
+
\widebar{\Omega}_{\mu,\bq}^\pm
\delta\mathcal{C}^{\zeta\zeta'}_\bq
+\frac{i\hbar}{2}\sqrt{2\gamma^{\rm x}}
\widebar{\Phi}^\pm_{\mu,0}
(\ev*{P^\dagger_{\zeta,0}}\delta P^{\rm in\dagger}_{\zeta',0}
+ \ev*{P^\dagger_{\zeta',0}}\delta P^{\rm in\dagger}_{\zeta,0})]
\\
-i\hbar\partial_t\delta{\mathcal{C}}^{\zeta\zeta'}_{\bq} &=
(\tilde{E}^{\rm x}_\bq + \tilde{E}^{\rm p}_\bq-2\hbar\omega_{\rm d})\delta{\mathcal{C}}^{\zeta\zeta'}_{\bq}
+ \Omega_\bq\delta\mathcal{D}^{\zeta\zeta'}_\bq
+ \sum_{\mu\pm} \Omega^\pm_{\mu,\bq} \delta\mathcal{B}^{\zeta\zeta'}_{\mu,\pm}
\\
&\hspace{2cm}+i\hbar\delta_{\bq,0}\qty[\sqrt{2\gamma^{\rm p}}
\qty(\ev*{a^{\rm in\dagger}_{\zeta,0}}\delta P^\dagger_{\zeta',0}
+\ev*{P^\dagger_{\zeta',0}}\delta a^{\rm in\dagger}_{\zeta,0})
+\sqrt{2\gamma^{\rm x}}\ev*{a^\dagger_{\zeta,0}}\delta P^{\rm in\dagger}_{\zeta',0}]
\\
-i\hbar\partial_t\delta\mathcal{D}^{\zeta\zeta'}_\bq &=
2(\tilde{E}^{\rm p}_\bq-\hbar\omega_{\rm d})\delta\mathcal{D}^{\zeta\zeta'}_\bq
+ \Omega_\bq\delta\mathcal{C}^{\zeta'\zeta}_{-\bq}
+ \Omega_{-\bq}\delta\mathcal{C}^{\zeta\zeta'}_{\bq}
\\ &\hspace{2cm}
+ i\hbar\sqrt{2\gamma^{\rm p}}\delta_{\bq,0}
\qty[
\ev*{a^{\rm in\dagger}_{\zeta}}\delta a^\dagger_{\zeta',0}
+ \ev*{a^{\rm in\dagger}_{\zeta'}}\delta a^\dagger_{\zeta,0}
+ \ev*{a^\dagger_{\zeta,0}}\delta a^{\rm in\dagger}_{\zeta'}
+ \ev*{a^\dagger_{\zeta',0}}\delta a^{\rm in\dagger}_{\zeta}
]
\end{split}
\end{align}
which are expressed in the rotating frame.

As described in the main text, we solve the fluctuation equations, Eqs.~\eqref{eq:photon-fluctuation-eom}, \eqref{eq:exciton-fluctuation-eom} and \eqref{eq:multiparticle-fluctuation-eom} in the steady-state limit, i.e. for $t\rightarrow\infty$, where the expectation values are constant. Here, we Fourier transform the Heisenberg-Langevin equations as $\delta Q(\omega) = \int_{-\infty}^{\infty}dt e^{i\omega t}\delta Q(t)$, $\delta Q$ being any of the fluctuation operators. The first step is to formally solve Eqs.~\eqref{eq:multiparticle-fluctuation-eom} in terms of the Langevin noises and $\delta a^\dagger_{\zeta,0}$ and $\delta P^\dagger_{\zeta,0}$.
The formal solutions of $\delta\mathcal{B}(\omega)$ and $\delta\mathcal{D}(\omega)$ are
\begin{align}
\label{eq:dB-dD-formal-solutions}
\begin{split}
  \delta\mathcal{B}_{\mu,\pm}^{\zeta\zeta'}(\omega)
  &= \frac{-1}{\hbar\omega+\tilde{E}^{\rm xx}_{\mu,\pm}-2\hbar\omega_{\rm d}}
  \frac{1}{2}(1\pm\delta_{\zeta\zeta'})
  \Big\{
  \sum_{\bq}\qty[
  \widebar{\Omega}_{\mu,-\bq}^\pm
  \delta\mathcal{C}^{\zeta'\zeta}_{-\bq}(\omega)
  +
  \widebar{\Omega}_{\mu,\bq}^\pm
  \delta\mathcal{C}^{\zeta\zeta'}_\bq(\omega)
  ]
  \\ &\hspace{5cm}+\frac{i\hbar}{2}\sqrt{2\gamma^{\rm x}}
  \widebar{\Phi}^\pm_{\mu,0}
  \qty[\ev*{P^\dagger_{\zeta,0}}\delta P^{\rm in\dagger}_{\zeta',0}(\omega)
  + \ev*{P^\dagger_{\zeta',0}}\delta P^{\rm in\dagger}_{\zeta,0}(\omega)]\Big\}
  \\
  \delta\mathcal{D}^{\zeta\zeta'}_\bq\!(\omega) &=
  \frac{-1}{\hbar\omega+2(\tilde{E}^{\rm p}_\bq-\hbar\omega_{\rm d})}
  \Big\{ \Omega_\bq\delta\mathcal{C}^{\zeta'\zeta}_{-\bq}\!(\omega)
  + \Omega_{-\bq}\delta\mathcal{C}^{\zeta\zeta'}_{\bq}\!(\omega)
  \\ &\hspace{2cm}
  + i\hbar\sqrt{2\gamma^{\rm p}}\delta_{\bq,0}
  \qty[
  \ev*{a^{\rm in\dagger}_{\zeta}}\delta a^\dagger_{\zeta',0}(\omega)
  + \ev*{a^{\rm in\dagger}_{\zeta'}}\delta a^\dagger_{\zeta,0}(\omega)
  + \ev*{a^\dagger_{\zeta,0}}\delta a^{\rm in\dagger}_{\zeta'}(\omega)
  + \ev*{a^\dagger_{\zeta',0}}\delta a^{\rm in\dagger}_{\zeta}(\omega)
  ]\Big\}
\end{split}
\end{align}
These are inserted into the equation for $\delta\mathcal{C}(\omega)$ and solved, thereby yielding
\begin{align}
\label{eq:dC-formal-solution}
  \begin{split}
    \delta\mathcal{C}^{\zeta\zeta'}_{\bq}\!\!(\omega) &=
    i\hbar\sqrt{2\gamma^{\rm p}}\sum_{\zeta_1\zeta_1'}
    K^{\zeta\zeta'\bq}_{\zeta_1\zeta_1'0}(\omega)
    [\ev*{a^{\rm in\dagger}_{\zeta_1}}\delta P^\dagger_{\zeta_1',0}(\omega)
    +
    \delta a^{\rm in\dagger}_{\zeta_1}(\omega)\ev*{P^\dagger_{\zeta_1',0}}]
    \\
    &-i\hbar\sqrt{2\gamma^{\rm p}}\sum_{\zeta_1\zeta_1'}
    K^{\zeta\zeta'\bq}_{\zeta_1\zeta_1'0}(\omega)
    \frac{\Omega_0}{\hbar\omega+2(\tilde{E}^{\rm p}_0-\hbar\omega_{\rm d})}
    \qty[
    \ev*{a^{\rm in\dagger}_{\zeta_1}}\delta a^\dagger_{\zeta_1',0}(\omega)
    + \ev*{a^{\rm in\dagger}_{\zeta_1'}}\delta a^\dagger_{\zeta_1,0}(\omega)]
    \\
    &-i\hbar\sqrt{2\gamma^{\rm p}}\sum_{\zeta_1\zeta_1'}
    K^{\zeta\zeta'\bq}_{\zeta_1\zeta_1'0}(\omega)
    \frac{\Omega_0}{\hbar\omega+2(\tilde{E}^{\rm p}_0-\hbar\omega_{\rm d})}
    \qty[
    \ev*{a^\dagger_{\zeta_1,0}}\delta a^{\rm in\dagger}_{\zeta_1'}(\omega)
    + \ev*{a^\dagger_{\zeta_1',0}}\delta a^{\rm in\dagger}_{\zeta_1}(\omega)]
    \\ &\hspace{-1cm}+ i\hbar\sqrt{2\gamma^{\rm x}}
    \sum_{\zeta_1\zeta_1'\bq_1}
    K^{\zeta\zeta'\bq}_{\zeta_1\zeta_1'\bq_1}\!\!(\omega)
    \qty{\delta_{\bq_1,0}\ev*{a^\dagger_{\zeta_1,0}}\delta P^{\rm in\dagger}_{\zeta_1',0}(\omega)-\sum_{\mu\pm}
    \frac{\frac{1}{4}(1\pm\delta_{\zeta_1\zeta_1'})\widebar{\Phi}^\pm_{\mu,0}\Omega^\pm_{\mu,\bq_1}}{\hbar\omega+\tilde{E}^{\rm xx}_{\mu,\pm}-2\hbar\omega_{\rm d}}
    \qty[\ev*{P^\dagger_{\zeta_1,0}}\delta P^{\rm in\dagger}_{\zeta_1',0}(\omega)
    + \ev*{P^\dagger_{\zeta_1',0}}\delta P^{\rm in\dagger}_{\zeta_1,0}(\omega)]},
  \end{split}
\end{align}
where
\begin{align}
K(\omega) = -\qty[
\delta_{\bq,\bq_1}\delta_{\zeta\zeta_1}\delta_{\zeta',\zeta_1'}
(\hbar\omega + \tilde{E}^{\rm x}_\bq + \tilde{E}^{\rm p}_\bq-2\hbar\omega_{\rm d})
+ \Pi^{\zeta\zeta'\bq}_{\zeta_1\zeta_1'\bq_1}(\omega)
]^{-1}.
\end{align}
is the Green's function for $\delta\mathcal{C}(\omega)$ with self-energy
\begin{align}
\begin{split}
  \Pi^{\zeta\zeta'\bq}_{\zeta_1\zeta_1'\bq_1}\!\!(\omega) &=
  -\frac{\Omega_\bq\Omega_{-\bq_1}}{\hbar\omega+2(\tilde{E}^{\rm p}_\bq-\hbar\omega_{\rm d})}
  \qty[\delta_{\zeta',\zeta_1}\delta_{\zeta,\zeta_1'}\delta_{-\bq,\bq_1}
  + \delta_{\zeta,\zeta_1}\delta_{\zeta',\zeta_1'}\delta_{\bq,\bq_1}]
  -\sum_{\mu\pm}
  \frac{\frac{1}{2}(1\pm\delta_{\zeta\zeta'})\Omega^\pm_{\mu,\bq} \widebar{\Omega}_{\mu,\bq_1}^\pm}{\hbar\omega+\tilde{E}^{\rm xx}_{\mu,\pm}-2\hbar\omega_{\rm d}}
  [\delta_{\zeta'\zeta_1}\delta_{\zeta,\zeta_1'}
  + \delta_{\zeta\zeta_1}\delta_{\zeta'\zeta_1'}].
\end{split}
\end{align}

We then insert the formal solution for $\delta\mathcal{B}$, Eq.~\eqref{eq:dB-dD-formal-solutions}, into Eq.~\eqref{eq:exciton-fluctuation-eom}, such that
\begin{align}
\label{eq:exciton-fluctuation-eom-2}
  \begin{split}
    -\hbar\omega\delta P_{\zeta,0}^\dagger(\omega)
    &= (\tilde{E}_0^{\rm x} - \hbar\omega_{\rm d}) \delta P_{\zeta,0}^\dagger(\omega)
    + \Omega_0\delta a_{\zeta,0}^\dagger(\omega)
    +\sum_{\zeta'}\Delta_{\zeta\zeta'}\delta P_{\zeta',0}(\omega)
    + i\hbar\sqrt{2\gamma^{\rm x}} \delta P_{\zeta,0}^{\rm in\dagger}(\omega),
    + \sum_{\zeta_1\zeta_2\bq} Q^\zeta_{\zeta_1\zeta_2\bq}(\omega)
    \delta\mathcal{C}^{\zeta_1\zeta_2}_\bq(\omega)
    \\
    &- i\hbar\sqrt{2\gamma^{\rm x}}
    \frac{1}{2}\sum_{\zeta'\mu\pm}
    \ev*{P_{\zeta',0}}
    \frac{\frac{1}{2}(1\pm\delta_{\zeta\zeta'})W^\pm_\mu\widebar{\Phi}^\pm_{\mu,0}}
    {\hbar\omega+\tilde{E}^{\rm xx}_{\mu,\pm}-2\hbar\omega_{\rm d}}
    \qty[\ev*{P^\dagger_{\zeta,0}}\delta P^{\rm in\dagger}_{\zeta',0}(\omega)
    + \ev*{P^\dagger_{\zeta',0}}\delta P^{\rm in\dagger}_{\zeta,0}(\omega)],
  \end{split}
\end{align}
where $\Delta_{\zeta,\zeta'}$ is defined in the main text and
\begin{align}
Q^\zeta_{\zeta_1\zeta_2\bq}\!(\omega) = -
\qty[
\delta_{\zeta_1\zeta}\delta_{\zeta_2\zeta}\tilde{\Omega}_\bq\ev*{P_{\zeta,0}}
+
\sum_{\mu\pm}
\frac{\frac{1}{2}(1\pm\delta_{\zeta_1\zeta_2})W^\pm_\mu\widebar{\Omega}_{\mu,\bq}^\pm
(\delta_{\zeta_2\zeta}\ev*{P_{\zeta_1,0}} + \delta_{\zeta_1\zeta}\ev*{P_{\zeta_2,0}})}
{\hbar\omega+\tilde{E}^{\rm xx}_{\mu,\pm}-2\hbar\omega_{\rm d}}
].
\end{align}
By substituting the formal solution for $\delta\mathcal{C}$, Eq.~\eqref{eq:dC-formal-solution}, into Eq.~\eqref{eq:exciton-fluctuation-eom-2}, we arrive at Eq.~(2) in the main text, with
\begin{align}
\begin{split}
  \Sigma_{\zeta\zeta'}(\omega) &= i\hbar\sqrt{2\gamma^{\rm p}}\sum_{\zeta_1\zeta_1'\bq_1}\sum_{\zeta_2\zeta_2'}
  Q^{\zeta}_{\zeta_1\zeta_1'\bq_1}\!\!(\omega)\;K^{\zeta_1\zeta_1'\bq_1}_{\zeta_2\zeta_2'0}(\omega)
  \ev*{a^{\rm in\dagger}_{\zeta_2}}\delta_{\zeta_2'\zeta'}
  \\
  \Omega_{\zeta\zeta',0}^{\mathrm{r}}(\omega) &=
  \Omega_0\delta_{\zeta,\zeta'}
  -i\hbar\sqrt{2\gamma^{\rm p}}\sum_{\zeta_1\zeta_1'\bq_1}\sum_{\zeta_2\zeta_2'}
  Q^{\zeta}_{\zeta_1\zeta_1'\bq_1}\!(\omega)
  K^{\zeta_1\zeta_1'\bq_1}_{\zeta_2\zeta_2'0}(\omega)
  \frac{\Omega_0}{\hbar\omega+2(\tilde{E}^{\rm p}_\bq-\hbar\omega_{\rm d})}
  \qty(\ev*{a^{\rm in\dagger}_{\zeta_2}}\delta_{\zeta',\zeta_2'}
  + \ev*{a^{\rm in\dagger}_{\zeta_2'}}\delta_{\zeta',\zeta_2})
  \\
  [\sqrt{2\Gamma^{\rm x}(\omega)}]_{\zeta\zeta'} &= \sqrt{2\gamma^{\rm x}}\Bigg\{\delta_{\zeta,\zeta'}
  - \frac{1}{2}\sum_{\zeta_1\mu\pm}
  \frac{\frac{1}{2}(1\pm\delta_{\zeta\zeta_1})
  W^\pm_\mu  \ev*{P_{\zeta_1,0}}
  \widebar{\Phi}^\pm_{\mu,0} }
  {\hbar\omega+\tilde{E}^{\rm xx}_{\mu,\pm}-2\hbar\omega_{\rm d}}
  \qty(\ev*{P^\dagger_{\zeta,0}}\delta_{\zeta',\zeta_1}
  + \ev*{P^\dagger_{\zeta_1,0}}\delta_{\zeta',\zeta})
  \\ &\hspace{-1.5cm}+ \sum_{\zeta_1\zeta_1'\bq_1}\sum_{\zeta_2\zeta_2'\bq_2}
    Q^\zeta_{\zeta_1\zeta_1'\bq_1}\!(\omega)
    K^{\zeta_1\zeta_1'\bq_1}_{\zeta_2\zeta_2'\bq_2}\!\!(\omega)
    \qty[\delta_{\bq_2,0}\ev*{a^\dagger_{\zeta_2,0}}\delta_{\zeta'\zeta_2'}-\sum_{\mu\pm}
    \frac{\frac{1}{4}(1\pm\delta_{\zeta_2\zeta_2'})\widebar{\Phi}^\pm_{\mu,0}\Omega^\pm_{\mu,\bq_1}}{\hbar\omega+\tilde{E}^{\rm xx}_{\mu,\pm}-2\hbar\omega_{\rm d}}
    \qty(\ev*{P^\dagger_{\zeta_2,0}}\delta_{\zeta'\zeta_2'}
    + \ev*{P^\dagger_{\zeta_2',0}}\delta_{\zeta',\zeta_2})]
    \Bigg\}
  \\
[\sqrt{2\Gamma^{\rm p}(\omega)}]_{\zeta\zeta'}  &=
\sqrt{2\gamma^{\rm p}}\sum_{\zeta_1\zeta_1'\bq_1}\sum_{\zeta_2\zeta_2'}
    Q^\zeta_{\zeta_1\zeta_1'\bq_1}\!(\omega)
    K^{\zeta_1\zeta_1'\bq_1}_{\zeta_2\zeta_2'0}\!(\omega)
    \qty[\ev*{P_{\zeta_2',0}^\dagger}\delta_{\zeta'\zeta_2} -
    \frac{\Omega_0}{\hbar\omega+2(\tilde{E}^{\rm p}_\bq-\hbar\omega_{\rm d})}
    \qty(\ev*{a^\dagger_{\zeta_2,0}}\delta_{\zeta'\zeta_2'}
    + \ev*{a^\dagger_{\zeta_2',0}}\delta_{\zeta',\zeta_2})].
\end{split}
\end{align}

\section{Numerical calculations}
\label{sec:methods}
The numerical calculations in the paper have been performed for atomically thin $\mathrm{MoS_2}$ encapsulated with hexagonal BN on both sides.

We use the screened Coulomb potential obtained from solving Poison's equation for the van der Waals heterostructure: dielectric environment/air gap/atomically thin semiconductor/air gap/dielectric environment \cite{florian2018dielectricSI,steinhoff2020dynamicalSI}.
The small interlayer air gaps (chosen $h_{int} = 0.3$~nm) take account of naturally occurring air gaps between the atomically thin semiconductor and its dielectric environment \cite{rooney2017observingSI} described by the dielectric constant $\varepsilon_e$.

The parameters for monolayer ${\rm MoS_2}$ are: layer thickness $d_{\rm 2D} = 0.626$~nm  \cite{rasmussen2015computationalSI}, single particle band gap ${\varepsilon}_g = 2.48$~eV \cite{rasmussen2015computationalSI}, effective electron mass $m_{\rm e} = 0.43~m_0$ \cite{kormanyos2015kSI}, effective hole mass $m_{\rm h} = 0.54~m_0$ \cite{kormanyos2015kSI}, valence-conduction band momentum matrix element $\gamma = 0.222$~eV~nm \cite{kormanyos2015kSI} and in-plane dielectric constant $\epsilon_\perp = 12.8$ \cite{kumar2012tunableSI}.

The phonon-induced dephasing rate $\gamma^\text{x}$ is calculated according to the methods given in Ref.~\cite{selig2016excitonicSI}, without self-consistent inclusion of radiative broadening, because this is contained in the interaction with the quantized electromagnetic field.

\end{document}